\documentclass[a4paper,11pt]{article}

\usepackage{microtype}

\usepackage[T1]{fontenc}
\usepackage[utf8]{inputenc}

\usepackage{amsfonts}
\usepackage{amsmath}
\usepackage{amssymb}
\usepackage{mathtools}
\usepackage{bm}
\usepackage{braket}
\renewcommand{\vec}[1]{\mathbf{#1}}
\newcommand{\uds}{\,\mathrm{d}}
\newcommand{\udds}{\,\mathrm{d}^2}

\newcommand{\pd}[2]{\frac{\partial #1}{\partial #2}}
\newcommand{\grad}{\bm{\nabla}}
\newcommand{\mat}[1]{\bm{\mathsf{#1}}}
\newcommand{\abs}[1]{\left\lvert#1\right\rvert}
\newcommand{\norm}[1]{\left\lVert#1\right\rVert}
\DeclareMathOperator{\diag}{diag}

\usepackage[pdftex]{graphicx}
\usepackage{siunitx}

\usepackage{caption}
\usepackage{subcaption}
\usepackage{txfonts}

\usepackage{multicol}
\usepackage{booktabs}

\makeatletter
\newcommand\tabfill[1]{%
\dimen@\linewidth%
\advance\dimen@\@totalleftmargin%
\advance\dimen@-\dimen\@curtab%
\parbox[t]\dimen@{#1\ifhmode\strut\fi}%
}

\newcommand{\ccol}[1]{\multicolumn{1}{c}{#1}}

\usepackage{color}
\definecolor{thered}{rgb}{0.65,0.04,0.07}
\definecolor{thegreen}{rgb}{0.06,0.44,0.08}
\definecolor{theblue}{rgb}{0.02,0.2,0.68}

\usepackage[numbers,sort]{natbib}
\usepackage[unicode=true,
            bookmarks=true,
            bookmarksnumbered=false,
            bookmarksopen=false,
            breaklinks=true,
            pdfborder={0 0 1},
            backref=false,
            colorlinks=true,
            linkcolor=black,
            urlcolor=theblue,
            citecolor=theblue,
            hyperfootnotes=false]
           {hyperref}
\hypersetup{pdftitle={PyFR},
            pdfauthor={F. D. Witherden, A. Farrington, P. E. Vincent}}

\begin{document}

\title{PyFR: An Open Source Framework for Solving Advection-Diffusion
  Type Problems on Streaming Architectures using the Flux Reconstruction
  Approach}

\author{F. D. Witherden\footnote{Corresponding author; e-mail freddie.witherden08@imperial.ac.uk.}, A. M. Farrington, P. E. Vincent\\\\
  \textit{\small Department of Aeronautics, Imperial College London, SW7 2AZ}}
\maketitle

\begin{abstract}
  High-order numerical methods for unstructured grids combine the
  superior accuracy of high-order spectral or finite difference methods
  with the geometric flexibility of low-order finite volume or finite
  element schemes. The Flux Reconstruction (FR) approach unifies various
  high-order schemes for unstructured grids within a single
  framework. Additionally, the FR approach exhibits a significant degree
  of element locality, and is thus able to run efficiently on modern
  streaming architectures, such as Graphical Processing Units
  (GPUs). The aforementioned properties of FR mean it offers a promising
  route to performing affordable, and hence industrially relevant,
  scale-resolving simulations of hitherto intractable unsteady flows
  within the vicinity of real-world engineering geometries.  In this
  paper we present PyFR, an open-source Python based framework for
  solving advection-diffusion type problems on streaming architectures
  using the FR approach. The framework is designed to solve a range of
  governing systems on mixed unstructured grids containing various
  element types.  It is also designed to target a range of hardware
  platforms via use of an in-built domain specific language based on the Mako templating engine. The current release of PyFR is able to solve the compressible Euler and
  Navier-Stokes equations on grids of quadrilateral and triangular
  elements in two dimensions, and hexahedral elements in three
  dimensions, targeting clusters of CPUs, and NVIDIA GPUs.  Results are
  presented for various benchmark flow problems, single-node performance is discussed, and scalability of the code is demonstrated on up to 104 NVIDIA M2090 GPUs. The software is freely available under a 3-Clause New Style BSD license (see www.pyfr.org).
\end{abstract}

{\footnotesize\emph{Keywords:} High-order; Flux reconstruction;
Parallel algorithms; Heterogeneous computing}

\newpage

\section*{Program Description}
\begin{description}
\item[Authors] F. D. Witherden, A. M. Farrington, P. E. Vincent
\item[Program title] PyFR v0.1.0
\item[Licensing provisions] New Style BSD license
\item[Programming language] Python, CUDA and C
\item[Computer] Variable, up to and including GPU clusters
\item[Operating system] Recent version of Linux/UNIX
\item[RAM] Variable, from hundreds of megabytes to gigabytes
\item[Number of processors used] Variable, code is multi-GPU and
  multi-CPU aware through a combination of MPI and OpenMP
\item[External routines/libraries] Python 2.7, numpy, PyCUDA, mpi4py,
  SymPy, Mako
\item[Nature of problem] Compressible Euler and Navier-Stokes equations
  of fluid dynamics; potential for any advection-diffusion type problem.
\item[Solution method] High-order flux reconstruction approach suitable
  for curved, mixed, unstructured grids.
\item[Unusual features] Code makes extensive use of symbolic
  manipulation and run-time code generation through a domain specific
  language.
\item[Running time] Many small problems can be solved on a recent
  workstation in minutes to hours.
\end{description}

\section*{Nomenclature}

Throughout we adopt a convention in which dummy indices on the right
hand side of an expression are summed.  For example $C_{ijk} =
A_{ijl}B_{ilk} \equiv \sum_l A_{ijl}B_{ilk}$ where the limits are
implied from the surrounding context.  All indices are assumed to be
zero-based.

\begin{multicols}{2}
\begin{tabbing}
\quad\qquad\= \kill
{\bfseries{Functions.}}\\
  $\delta_{ij}$ \> Kronecker delta\\
  $\det \mat{A}$ \> Matrix determinant\\
  $\dim \mat{A}$ \> Matrix dimensions\\\\
{\bfseries{Indices.}}\\
  $e$ \> Element type\\
  $n$ \> Element number\\
  $\alpha$ \> Field variable number\\
  $i,j,k$ \> Summation indices\\
  $\rho,\sigma,\nu$ \> Summation indices\\\\
{\bfseries{Domains.}}\\
  $\vec{\Omega}$ \> Solution domain\\
  $\vec{\Omega}_e$ \> All elements in $\vec{\Omega}$ of type $e$\\
  $\hat{\vec{\Omega}}_{e}$ \> A \emph{standard} element of type $e$\\
  $\partial\hat{\vec{\Omega}}_{e}$ \> Boundary of $\hat{\vec{\Omega}}_{e}$\\
  $\vec{\Omega}_{en}$ \> \tabfill{Element $n$ of type $e$ in
  $\vec{\Omega}$}\\
  $\abs{\vec{\Omega}_e}$ \> Number of elements of type $e$\\\\
{\bfseries{Expansions.}}\\
  $\wp$ \> Polynomial order\\
  $N_D$ \> Number of spatial dimensions\\
  $N_V$ \> Number of field variables\\
  $\ell_{e\rho}$ \> \tabfill{Nodal basis polynomial $\rho$ for element type $e$}\\
  $x,y,z$ \> Physical coordinates\\
  $\tilde{x},\tilde{y},\tilde{z}$ \> Transformed coordinates\\
  $\bm{\mathcal{M}}_{en}$ \> Transformed to physical mapping\\\\
{\bfseries{Adornments and suffixes.}}\\
  $\tilde{\square}$ \> \tabfill{A quantity in transformed space}\\
  $\hat{\square}$ \> \tabfill{A vector quantity of unit magnitude}\\
  $\square^T$ \> Transpose\\
  $\square^{(u)}$ \> A quantity at a solution point\\
  $\square^{(f)}$ \> A quantity at a flux point\\
  $\square^{(f_\perp)}$ \> \tabfill{A normal quantity at a flux point}\\\\
{\bfseries{Operators.}}\\
  $\mathfrak{C}_\alpha$ \> \tabfill{Common solution at an interface}\\
  $\mathfrak{F}_\alpha$ \> \tabfill{Common normal flux at an interface}\\\\
\end{tabbing}
\end{multicols}

\section{Introduction}

There is an increasing desire amongst industrial practitioners of
computational fluid dynamics (CFD) to undertake high-fidelity
scale-resolving simulations of transient compressible flows within the
vicinity of complex geometries.  For example, to improve the design of
next generation unmanned aerial vehicles (UAVs), there exists a need to
perform simulations---at Reynolds numbers $10^4$--$10^7$ and Mach
numbers $M\sim 0.1$--$1.0$---of highly separated flow over deployed
spoilers/air-brakes; separated flow within serpentine intake ducts;
acoustic loading in weapons bays; and flow over entire UAV
configurations at off-design conditions.  Unfortunately, current
generation industry-standard CFD software based on first- or
second-order accurate Reynolds Averaged Navier-Stokes (RANS) approaches
is not well suited to performing such simulations.  Henceforth, there
has been significant interest in the potential of high-order accurate
methods for unstructured mixed grids, and whether they can offer an
efficient route to performing scale-resolving simulations within the
vicinity of complex geometries. Popular examples of high-order schemes
for unstructured mixed grids include the discontinuous Galerkin (DG)
method, first introduced by Reed and Hill \cite{reed1973triangular}, and
the spectral difference (SD) methods originally proposed under the
moniker `staggered-gird Chebyshev multidomain methods' by Kopriva and
Kolias in 1996 \cite{kopriva1996conservative} and later popularised by
Sun et al. \cite{sun2007high}. In 2007 Huynh proposed the flux
reconstruction (FR) approach \cite{huynh2007flux}; a unifying framework
for high-order schemes for unstructured grids that incorporates both the
nodal DG schemes of \cite{hesthaven2008nodal} and, at least for a linear
flux function, any SD scheme.  In addition to offering high-order
accuracy on unstructured mixed grids, FR schemes are also compact in
space, and thus when combined with explicit time marching offer a
significant degree of element locality. As such, explicit high-order FR
schemes are characterised by a large degree of structured computation.

\begin{figure}
  \centering
  \includegraphics{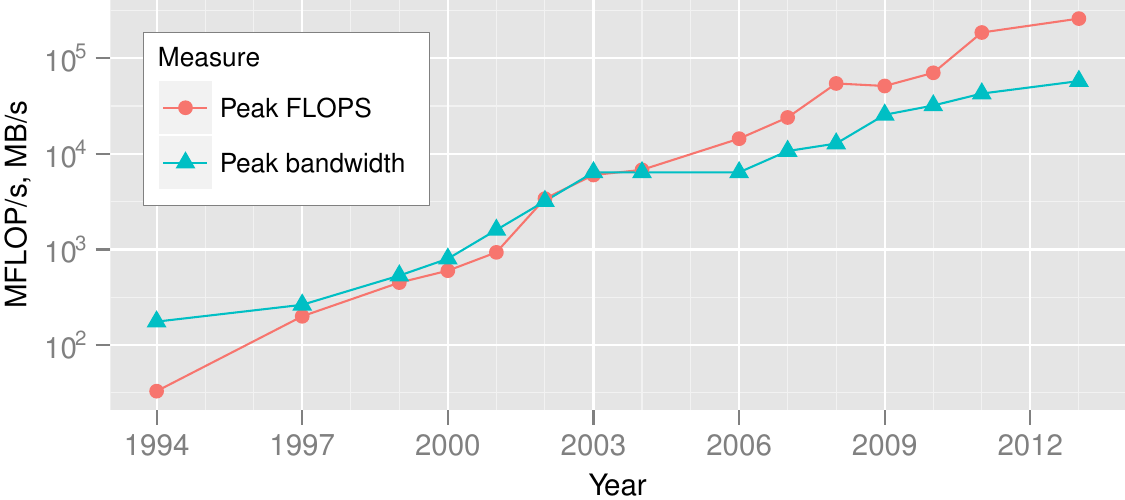}%
  \caption{\label{fig:memwall}Trends in the peak floating point
    performance (double precision) and memory bandwidth of sever-class
    Intel processors from 1994--2013.  The quotient of these two
    measures yields the FLOPS-per-byte of a processor.  Data courtesy
    of Jan Treibig.}
\end{figure}

Over the past two decades improvements in the arithmetic capabilities of
processors have significantly outpaced advances in random access memory.
Algorithms which have traditionally been compute bound---such as dense
matrix-vector products---are now limited instead by the bandwidth
to/from memory.  This is epitomised in \autoref{fig:memwall}.  Whereas
the processors of two decades ago had FLOPS-per-byte of ${\sim}0.2$ more
recent chips have ratios upwards of ${\sim}4$.  This disparity is not
limited to just conventional CPUs.  Massively parallel accelerators and
co-processors such as the NVIDIA K20X and Intel Xeon Phi 5110P have
ratios of $5.24$ and $3.16$, respectively.

A concomitant of this disparity is that modern hardware architectures
are highly dependent on a combination of high speed caches and/or shared
memory to maintain throughput.  However, for an algorithm to utilise
these efficiently its memory access pattern must exhibit a degree of
either spatial or temporal locality.  To a first-order approximation the
spatial locality of a method is inversely proportional to the amount of
memory indirection. On an unstructured grid indirection arises whenever
there is coupling between elements.  This is potentially a problem for
discretisations whose stencil is not compact.  Coupling also arises in
the context of implicit time stepping schemes.  Implementations are
therefore very often bound by memory bandwidth.  As a secondary trend we
note that the manner in which FLOPS are realised has also changed.  In
the early 1990s commodity CPUs were predominantly scalar with a single
core of execution.  However in 2013 processors with eight or more cores
are not uncommon.  Moreover, the cores on modern processors almost
always contain vector processing units.  Vector lengths up to 256-bits,
which permit up to four double precision values to be operated on at
once, are not uncommon.  It is therefore imperative that compute-bound
algorithms are amenable to both multithreading and vectorisation.  A
versatile means of accomplishing this is by breaking the computation
down into multiple, necessarily independent, streams.  By virtue of
their independence these streams can be readily divided up between cores
and vector lanes.  This leads directly to the concept of \emph{stream
  processing}.  We will refer to architectures amenable to this form of
parallelisation as streaming architectures.

A corollary of the above discussion is that compute intensive
discretisations which can be formulated within the stream processing
paradigm are well suited to acceleration on current---and likely
future---hardware platforms.  The FR approach combined with explicit
time stepping is an archetypical of this.

Our objective in this paper is to present PyFR, an open-source Python
based framework for solving advection-diffusion type problems on
streaming architectures using the FR approach. The framework is designed
to solve a range of governing systems on mixed unstructured grids
containing various element types. It is also designed to target a range
of hardware platforms via use of an in-built domain specific language
derived from the Mako templating engine. The current release of PyFR is
able to solve the compressible Euler and Navier-Stokes equations on
unstructured grids of quadrilateral and triangular elements in
two-dimensions, and unstructured grids of hexahedral elements in
three-dimensions, targeting clusters of CPUs, and NVIDIA GPUs. The paper
is structured as follows. In \autoref{sec:discretisation} we provide a
overview of the FR approach for advection-diffusion type
problems on mixed unstructured grids. In
\autoref{sec:implementation} we proceed to describe our implementation
strategy, and in \autoref{sec:govsystems} we present the Euler and
Navier-Stokes equations, which are solved by the current release of
PyFR. The framework is then validated in \autoref{sec:validation}, single-node performance is discussed in \autoref{sec:performance}, and scalability of the code is demonstrated on up to 104 NVIDIA M2090 GPUs in \autoref{sec:scalability}. Finally, conclusions are drawn in
\autoref{sec:conclusion}.

\section{Flux Reconstruction}
\label{sec:discretisation}

A brief overview of the FR approach for solving advection-diffusion type problems is given below. Extended presentations can be found elsewhere \cite{huynh2007flux,vincent2011new,castonguay2011new,castonguay2011application,jameson2011non,vincent2011insights,castonguay2012new,williams2013energyb,castonguay2013energy,williams2013tet}.

Consider the following advection-diffusion problem inside an arbitrary
domain $\vec{\Omega}$ in $N_D$ dimensions
\begin{equation} \label{eq:advdiffpde}
  \pd{u_\alpha}{t} + \grad \cdot \vec{f}_\alpha = 0,
\end{equation}
where $0 \le \alpha < N_V$ is the \emph{field variable} index, $u_\alpha
= u_\alpha(\vec{x},t)$ is a conserved quantity, $\vec{f}_\alpha =
\vec{f}_\alpha(u, \grad u)$ is the flux of this conserved quantity and
$\vec{x} = x_i \in \mathbb{R}^{N_D}$.  In defining the flux we have
taken $u$ in its unscripted form to refer to all of the $N_V$ field
variables and $\grad u$ to be an object of length $N_D \times N_V$
consisting of the gradient of each field variable.  We start by
rewriting \autoref{eq:advdiffpde} as a first order system
\begin{subequations}
\begin{align}
  \pd{u_\alpha}{t} + \grad \cdot \vec{f}_\alpha(u, \vec{q}) &= 0,\\
  \vec{q}_\alpha - \grad u_\alpha &= 0,
\end{align}
\end{subequations}
where $\vec{q}$ is an auxiliary variable.  Here, as with $\grad u$, we
have taken $\vec{q}$ in its unsubscripted form to refer to the gradients
of all of the field variables.

Take $\mathcal{E}$ to be the set of available element types in
$\mathbb{R}^{N_D}$.  Examples include quadrilaterals and triangles in two
dimensions and hexahedra, prisms, pyramids and tetrahedra in three
dimensions.  Consider using these various elements types to construct a
conformal mesh of the domain such that
\[
\vec{\Omega} = \bigcup_{e\in\mathcal{E}} \vec{\Omega}_e \qquad
\text{and} \qquad \vec{\Omega}_e = \bigcup_{n=0}^{|\vec{\Omega}_e|-1}
\vec{\Omega}_{en} \qquad \text{and} \qquad
\bigcap_{e\in\mathcal{E}}\bigcap_{n=0}^{|\vec{\Omega}_e|-1}
\vec{\Omega}_{en} = \emptyset,
\]
where $\vec{\Omega}_e$ refers to all of the elements of type $e$ inside
of the domain, $\abs{\vec{\Omega}_e}$ is the number of elements of this
type in the decomposition, and $n$ is an index running over these
elements with $0 \leq n < \abs{\vec{\Omega}_e}$.  Inside each element
$\vec{\Omega}_{en}$ we require that
\begin{subequations}
\begin{align}
  \pd{u_{en\alpha}}{t} +  \grad \cdot \vec{f}_{en\alpha} &= 0,\\
  \vec{q}_{en\alpha} - \grad u_{en\alpha} &= 0.
\end{align}
\end{subequations}

It is convenient, for reasons of both mathematical simplicity and
computational efficiency, to work in a transformed space.  We accomplish
this by introducing, for each element type, a standard element
$\vec{\hat{\Omega}}_{e}$ which exists in a transformed space,
$\tilde{\vec{x}} = \tilde{x}_i$.  Next, assume the existence of a
mapping function for each element such that
\begin{align*}
  x^{}_i &= \mathcal{M}_{eni}(\tilde{\vec{x}}), & \vec{x} &=
  \bm{\mathcal{M}}_{en}(\tilde{\vec{x}}), \\
  \tilde{x}^{}_i &= \mathcal{M}^{-1}_{eni}(\vec{x}), & \tilde{\vec{x}}
  &= \bm{\mathcal{M}}^{-1}_{en}(\vec{x}),
\end{align*}
along with the relevant Jacobian matrices
\begin{align*}
  \mat{J}_{en} = J_{enij} &= \pd{\mathcal{M}_{eni}}{\tilde{x}_j}, &
  J_{en}
  &= \det\mat{J}_{en},\\
  \mat{J}^{-1}_{en} = J^{-1}_{enij} &= \pd{\mathcal{M}^{-1}_{eni}}{x_j},
  & J^{-1}_{en} &= \det\mat{J}^{-1}_{en} = \frac{1}{J_{en}}.
\end{align*}
These definitions provide us with a means of transforming quantities to
and from standard element space.  Taking the transformed solution, flux,
and gradients inside each element to be
\begin{subequations}
\begin{align}
  \tilde{u}^{}_{en\alpha} &= \tilde{u}^{}_{en\alpha}(\tilde{\vec{x}},t)
  =
  J^{}_{en}(\tilde{\vec{x}})u^{}_{en\alpha}(\bm{\mathcal{M}}^{}_{en}(\tilde{\vec{x}}),t),\\
  \tilde{\vec{f}}^{}_{en\alpha} &=
  \tilde{\vec{f}}^{}_{en\alpha}(\tilde{\vec{x}},t) =
  J^{}_{en}(\tilde{\vec{x}})\mat{J}^{-1}_{en}(\bm{\mathcal{M}}^{}_{en}(\tilde{\vec{x}}))
  \vec{f}^{}_{en\alpha}(\bm{\mathcal{M}}^{}_{en}(\tilde{\vec{x}}),t),\\
  \tilde{\vec{q}}^{}_{en\alpha} &= \tilde{\vec{q}}^{}_{en\alpha}(\tilde{\vec{x}},t)
  = \mat{J}^{T}_{en}(\tilde{\vec{x}})\vec{q}^{}_{en\alpha}(\bm{\mathcal{M}}^{}_{en}(\tilde{\vec{x}}),t),
\end{align}
\end{subequations}
and letting $\tilde{\grad} = \partial/\partial\tilde{x}_i$, it can be
readily verified that
\begin{subequations}
  \begin{align} \pd{u_{en\alpha}}{t} +
    J^{-1}_{en}\tilde{\grad} \cdot \tilde{\vec{f}}_{en\alpha} &= 0,\\
    \tilde{\vec{q}}_{en\alpha} - \tilde{\grad} u_{en\alpha} &=
    0,\label{eq:advecpdetaux}
\end{align}
\end{subequations}
as required.  We note here the decision to multiply the first equation
through by a factor of $J^{-1}_{en}$.  Doing so has the effect of taking
$\tilde{u}_{en} \mapsto u_{en}$ which allows us to work in terms of the
physical solution.  This is more convenient from a computational
standpoint.

\begin{figure}
  \centering
  \includegraphics{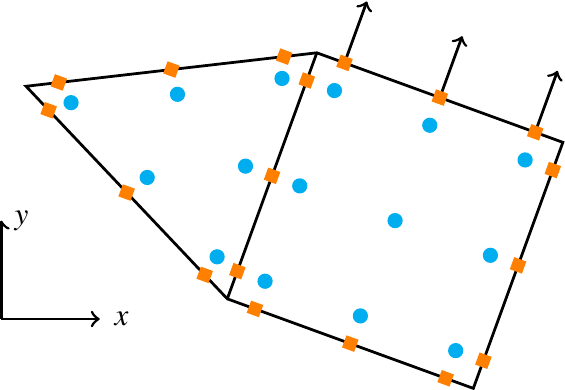}%
  \caption{\label{fig:elepts}Solution points (blue circles) and flux
    points (orange squares) for a triangle and quadrangle in physical
    space.  For the top edge of the quadrangle the normal vectors have
    been plotted.  Observe how the flux points at the interface between
    the two elements are co-located.}
\end{figure}

We next proceed to associate a set of
solution points with each standard element.  For each type $e \in \mathcal{E}$ take
$\set{\tilde{\vec{x}}^{(u)}_{e\rho}}$ to be the chosen set of points
where $0 \le \rho < N^{(u)}_e(\wp)$.  These points can then be used to
construct a nodal basis set $\set{\ell^{(u)}_{e\rho}(\tilde{\vec{x}})}$
with the property that
$\ell^{(u)}_{e\rho}(\tilde{\vec{x}}^{(u)}_{e\sigma}) =
\delta_{\rho\sigma}$.  To obtain such a set we first take
$\set{\psi_{e\sigma}(\tilde{\vec{x}})}$ to be any basis which spans a selected
order $\wp$ polynomial space defined inside $\hat{\vec{\Omega}}_e$.
Next we compute the elements of the generalised Vandermonde matrix
$\mathcal{V}_{e\rho\sigma} =
\psi_{e\rho}(\tilde{\vec{x}}^{(u)}_{e\sigma})$.  With these a nodal
basis set can be constructed as $\ell^{(u)}_{e\rho}(\tilde{\vec{x}}) =
\mathcal{V}^{-1}_{e\rho\sigma}\psi^{}_{e\sigma}(\tilde{\vec{x}})$.
Along with the solution points inside of each element we also define a
set of flux points on $\partial\hat{\vec{\Omega}}_e$.  We denote the
flux points for a particular element type as
$\set{\tilde{\vec{x}}^{(f)}_{e\rho}}$ where $0 \le \rho <
N^{(f)}_e(\wp)$.  Let the set of corresponding normalised
outward-pointing normal vectors be given by
$\set{\hat{\tilde{\vec{n}}}^{(f)}_{e\rho}}$.  It is critical that each
flux point pair along an interface share the same coordinates in
physical space.  For a pair of flux points $e\rho n$ and $e^\prime
\rho^\prime n^\prime$ at a non-periodic interface this can be formalised
as $\bm{\mathcal{M}}^{}_{en}(\tilde{\vec{x}}^{(f)}_{e\rho}) =
\bm{\mathcal{M}}^{}_{e^\prime n^\prime}(\tilde{\vec{x}}^{(f)}_{e^\prime
  \rho^\prime})$. A pictorial illustration of this can be seen in
\autoref{fig:elepts}.

The first step in the FR approach is to go from the discontinuous
solution at the solution points to the discontinuous solution at the
flux points
\begin{equation} \label{eq:ufpts}
  u^{(f)}_{e\sigma n\alpha} = u^{(u)}_{e\rho n\alpha}
  \ell^{(u)}_{e\rho}(\tilde{\vec{x}}^{(f)}_{e\sigma}),
\end{equation}
where $u^{(u)}_{e\rho n\alpha}$ is an approximate solution of field
variable $\alpha$ inside of the $n$th element of type $e$ at solution
point $\tilde{\vec{x}}^{(u)}_{e\rho}$.  This can then be used to compute
a \emph{common solution}
\begin{equation} \label{eq:commsoln}
  \mathfrak{C}^{\vphantom{(f)}}_{\vphantom{\widetilde{e\rho n}\alpha}\alpha}u^{(f)}_{\vphantom{\widetilde{e\rho n}\alpha}e\rho n\alpha}
  = \mathfrak{C}^{\vphantom{(f)}}_{\vphantom{\widetilde{e\rho n}\alpha}\alpha}u^{(f)}_{\widetilde{e\rho n}\alpha} =
  \mathfrak{C}^{\vphantom{(f)}}_{\vphantom{\widetilde{e\rho n}\alpha}\alpha}(u^{(f)}_{\vphantom{\widetilde{e\rho n}\alpha}e\rho n\alpha},
  u^{(f)}_{\widetilde{e\rho n}\alpha}),
\end{equation}
where $\mathfrak{C}_{\alpha}(u_L,u_R)$ is a scalar function that
given two values at a point returns a common value.  Here we have taken
$\widetilde{e\rho n}$ to be the element type, flux point number and
element number of the adjoining point at the interface.  Since grids in
FR are permitted to be unstructured the relationship between $e\rho n$
and $\widetilde{e\rho n}$ is indirect.  This necessitates the use of a
lookup table.  As the common solution function is permitted to perform
upwinding or downwinding of the solution it is in general the case that
$\mathfrak{C}^{\vphantom{(f)}}_{\vphantom{\widetilde{e\rho
      n}\alpha}\alpha}(u^{(f)}_{\vphantom{\widetilde{e\rho
      n}\alpha}e\rho n\alpha}, u^{(f)}_{\widetilde{e\rho n}\alpha}) \neq
\mathfrak{C}^{\vphantom{(f)}}_{\vphantom{\widetilde{e\rho
      n}\alpha}\alpha}(u^{(f)}_{\widetilde{e\rho n}\alpha},
u^{(f)}_{\vphantom{\widetilde{e\rho n}\alpha}e\rho n\alpha}) $. Hence,
it is important that each flux point pair only be visited \emph{once}
with the same common solution value assigned to both
$\mathfrak{C}^{\vphantom{(f)}}_{\vphantom{\widetilde{e\rho
      n}\alpha}\alpha}u^{(f)}_{\vphantom{\widetilde{e\rho n}\alpha}e\rho
  n\alpha}$ and
$\mathfrak{C}^{\vphantom{(f)}}_{\vphantom{\widetilde{e\rho
      n}\alpha}\alpha}u^{(f)}_{\widetilde{e\rho n}\alpha}$.

Further, associated with each flux point is a vector correction function
$\vec{g}^{(f)}_{e\rho}(\tilde{\vec{x}})$ constrained such that
\begin{equation}\label{eq:corfn}
\hat{\tilde{\vec{n}}}^{(f)}_{e\sigma} \cdot
\vec{g}^{(f)}_{e\rho}(\tilde{\vec{x}}^{(f)}_{e\sigma}) =
\delta^{}_{\rho\sigma},
\end{equation}
with a divergence that sits in the same
polynomial space as the solution.  Using these fields we can express the
solution to \autoref{eq:advecpdetaux} as
\begin{equation} \label{eq:transqu} \tilde{\vec{q}}^{(u)}_{e\sigma
    n\alpha} = \bigg[\hat{\tilde{\vec{n}}}^{(f)}_{e\rho}\cdot
  \tilde{\grad} \cdot
  \vec{g}^{(f)}_{e\rho}(\tilde{\vec{x}})\left\{\mathfrak{C}^{\vphantom{(f)}}_{\alpha}u^{(f)}_{e\rho
      n\alpha} - u^{(f)}_{e\rho n\alpha}\right\} + u^{(u)}_{e\nu
    n\alpha}
  \tilde{\grad}\ell^{(u)}_{e\nu}(\tilde{\vec{x}})\bigg]_{\tilde{\vec{x}}
    = \tilde{\vec{x}}^{(u)}_{e\sigma}},
\end{equation}
where the term inside the curly brackets is the `jump' at the interface
and the final term is an order $\wp - 1$ approximation of the gradient
obtained by differentiating the discontinuous solution polynomial.
Following the approaches of Kopriva \cite{kopriva1998staggered} and Sun
et al. \cite{sun2007high} we can now compute physical gradients as
\begin{align}
  \vec{q}^{(u)}_{e\sigma n\alpha} &= \mat{J}^{-T\,(u)}_{e\sigma n}\tilde{\vec{q}}^{(u)}_{e\sigma n\alpha},\\
  \vec{q}^{(f)}_{e\sigma n\alpha} &=
  \ell^{(u)}_{e\rho}(\tilde{\vec{x}}^{(f)}_{e\sigma})\vec{q}^{(u)}_{e\rho
    n\alpha},\label{eq:gradufpts}
\end{align}
where $\mat{J}^{-T\,(u)}_{e\sigma n} =
\mat{J}^{-T}_{en}(\tilde{\vec{x}}^{(u)}_{e\sigma})$.  Having solved the
auxiliary equation we are now able to evaluate the transformed flux
\begin{equation}
 \tilde{\vec{f}}^{(u)}_{e\rho n\alpha} = J^{(u)}_{e\rho n}\mat{J}^{-1\,(u)}_{e\rho n}
 \vec{f}^{}_{\alpha}(u^{(u)}_{e\rho n}, \vec{q}^{(u)}_{e\rho n}),
\end{equation}
where $J^{(u)}_{e\rho n} = \det
\mat{J}_{en}(\tilde{\vec{x}}^{(u)}_{e\rho})$.  This can be seen to be a
collocation projection of the flux.  With this it is possible to compute
the normal transformed flux at each of the flux points
\begin{equation} \label{eq:tnormfluxf}
  \tilde{f}^{(f_\perp)}_{e\sigma n\alpha} =
  \ell^{(u)}_{e\rho}(\tilde{\vec{x}}^{(f)}_{e\sigma}) \hat{\tilde{\vec{n}}}^{(f)}_{e\sigma} \cdot \tilde{\vec{f}}^{(u)}_{e\rho n\alpha}.
\end{equation}
Considering the physical normals at the flux points we see that
\begin{equation}
  \vec{n}^{(f)}_{e\sigma n} =  n^{(f)}_{e\sigma n}
  \hat{\vec{n}}^{(f)}_{e\sigma n} =  \mat{J}^{-T\,(f)}_{e\sigma n}\hat{\tilde{\vec{n}}}^{(f)}_{e\sigma},
\end{equation}
which is the outward facing normal vector in physical space where
$n^{(f)}_{e\sigma n} > 0$ is defined as the magnitude.  As the
interfaces between two elements conform we must have
$\hat{\vec{n}}^{(f)}_{\vphantom{\widetilde{e\sigma n}}e\sigma n} =
-\hat{\vec{n}}^{(f)}_{\widetilde{e\sigma n}}$.  With these definitions
we are now in a position to specify an expression for the \emph{common
  normal flux} at a flux point pair as
\begin{equation} \label{eq:commpflux}
\mathfrak{F}^{\vphantom{(f)}}_{\vphantom{\widetilde{e\sigma n}}\alpha}f^{(f_\perp)}_{\vphantom{\widetilde{e\sigma n}}e\sigma n\alpha} =
- \mathfrak{F}^{\vphantom{(f)}}_{\vphantom{\widetilde{e\sigma n}}\alpha}f^{(f_\perp)}_{\widetilde{e\sigma n}\alpha} =
 \mathfrak{F}^{\vphantom{(f)}}_{\vphantom{\widetilde{e\sigma n}}\alpha}(u^{(f)}_{\vphantom{\widetilde{e\sigma n}}e\sigma n},
u^{(f)}_{\widetilde{e\sigma n}},
\vec{q}^{(f)}_{\vphantom{\widetilde{e\sigma n}}e\sigma n},
\vec{q}^{(f)}_{\widetilde{e\sigma n}},
\hat{\vec{n}}^{(f)}_{\vphantom{\widetilde{e\sigma n}}e\sigma n}).
\end{equation}
The relationship
$\mathfrak{F}^{\vphantom{(f)}}_{\vphantom{\widetilde{e\sigma
      n}}\alpha}f^{(f_\perp)}_{\vphantom{\widetilde{e\sigma n}}e\sigma
  n\alpha} =
-\mathfrak{F}^{\vphantom{(f)}}_{\vphantom{\widetilde{e\sigma
      n}}\alpha}f^{(f_\perp)}_{\widetilde{e\sigma n}\alpha}$ arises from
the desire for the resulting numerical scheme to be conservative; a net
outward flux from one element must be balanced by a corresponding inward
flux on the adjoining element.  It follows that that
$\mathfrak{F}_{\alpha}(u_L, u_R, \vec{q}_L, \vec{q}_R, \hat{\vec{n}}_L)
= -\mathfrak{F}_{\alpha}(u_R, u_L, \vec{q}_R, \vec{q}_L,
-\hat{\vec{n}}_L)$.  The common normal fluxes in \autoref{eq:commpflux}
can now be taken into transformed space via
\begin{align}
 \mathfrak{F}^{\vphantom{(f_\perp)}}_{\alpha}\tilde{f}^{(f_\perp)}_{e\sigma n\alpha} &=
 J^{(f)}_{e\sigma n}n^{(f)}_{e\sigma n}\mathfrak{F}^{}_{\alpha}f^{(f_\perp)}_{e\sigma n\alpha},
 \\
 \mathfrak{F}^{\vphantom{(f_\perp)}}_{\vphantom{\widetilde{e\sigma n}}\alpha}\tilde{f}^{(f_\perp)}_{\widetilde{e\sigma n}\alpha}
 &= J^{(f)}_{\widetilde{e\sigma n}}n^{(f)}_{\widetilde{e\sigma n}}\mathfrak{F}^{\vphantom{(f_\perp)}}_{\vphantom{\widetilde{e\sigma n}}\alpha}f^{(f_\perp)}_{\widetilde{e\sigma n}\alpha},
\end{align}
where $J^{(f)}_{e\sigma n} = \det
\mat{J}_{en}(\tilde{\vec{x}}^{(f)}_{e\sigma})$.

It is now possible to compute an approximation for the divergence of the
\emph{continuous} flux.  The procedure is directly analogous to the one
used to calculate the transformed gradient in \autoref{eq:transqu}
\begin{equation} \label{eq:tdivflux}
  (\tilde{\grad} \cdot \tilde{\vec{f}})^{(u)}_{e\rho n\alpha} =
  \bigg[\tilde{\grad} \cdot \vec{g}^{(f)}_{e\sigma}(\tilde{\vec{x}})
  \left\{\mathfrak{F}^{\vphantom{(f)}}_{\alpha}\tilde{f}^{(f_\perp)}_{e\sigma n\alpha} -
    \tilde{f}^{(f_\perp)}_{e\sigma n\alpha}\right\} +
  \tilde{\vec{f}}^{(u)}_{e\nu n\alpha} \cdot \tilde{\grad}\ell^{(u)}_{e\nu}(\tilde{\vec{x}})\bigg]_{\tilde{\vec{x}} =
    \tilde{\vec{x}}^{(u)}_{e\rho}},
\end{equation}
which can then be used to obtain a semi-discretised form of the
governing system
\begin{equation} \label{eq:negdivconf}
  \pd{u^{(u)}_{e\rho n\alpha}}{t} =
  -J^{-1\,(u)}_{e\rho n}(\tilde{\grad} \cdot
    \tilde{\vec{f}})^{(u)}_{e\rho n\alpha},
\end{equation}
where $J^{-1\,(u)}_{e\rho n} = \det
\mat{J}^{-1}_{en}(\tilde{\vec{x}}^{(u)}_{e\rho}) = 1/J^{(u)}_{e\rho n}$.

This semi-discretised form is simply a system of ordinary differential
equations in $t$ and can be solved using one of a number of schemes,
e.g. a classical fourth order Runge-Kutta (RK4) scheme.

\section{Implementation}
\label{sec:implementation}

\subsection{Overview}

PyFR is a Python based implementation of the FR approach described in section \autoref{sec:discretisation}. It is designed to be compact, efficient, and platform portable. Key functionality is summarised in table \autoref{tab:pyfr-func}.

\begin{table}
  \centering
  \caption{\label{tab:pyfr-func}Key functionality of PyFR.}
  \begin{tabular}{rl} \toprule
    Dimensions     & 2D, 3D \\
    Elements       & Triangles, Quadrilaterals, Hexahedra\\
    Spatial orders & Arbitrary\\
    Time steppers  & Euler, RK4, DOPRI5\\
    Precisions     & Single, Double\\
    Platforms       & CPUs via C/OpenMP, Nvidia GPUs via CUDA\\
    Communication  & MPI\\
    Governing Systems  & Euler, Compressible Navier-Stokes\\
    \bottomrule
  \end{tabular}
\end{table}

The majority of operations within an FR step can be cast in terms of
matrix-matrix multiplications, as detailed in \autoref{sec:mat-rep}. All
remaining operations (\textit{e.g.} flux evaluations) are point-wise,
concerning themselves with either a single solution point inside of an
element or two collocating flux points at an interface. Hence, in broad
terms, there are five salient aspects of an FR implementation,
specifically i.) definition of the constant operator matrices detailed
in \autoref{sec:mat-rep}, ii.) specification of the state matrices
detailed in \autoref{sec:mat-rep}, iii.) implementation of matrix
multiply kernels, iv.) implementation of point-wise kernels, and finally
v.) handling of distributed memory parallelism and scheduling of kernel
invocations. Details regarding how each of the above were achieved in
PyFR are presented below.

\subsection{Definition of Constant Operator Matrices}

Setup of the seven constant operator matrices detailed in
\autoref{sec:mat-rep} requires evaluation of various polynomial
expressions, and their derivatives, at solution/flux points within each
type of standard element. Although conceptually simple, such operations
can be cumbersome to code. To keep the codebase compact PyFR makes
extensive use of symbolic manipulation via SymPy \cite{sympy2013}, which
brings computer algebra facilities similar to those found in Maple and
Mathematica to Python. SymPy has built-in support for most common
polynomials and can readily evaluate such expressions to arbitrary
precision. Efficiency of the setup phase is not critical, since the
operations are only performed once at start-up. Since efficiency is not
critical, platform portability is effectively achieved by running such
operations on the host CPU in all cases.

\subsection{Specification of State Matrices}
\label{sec:layout}

In specifying the state matrices detailed in \autoref{sec:mat-rep} there is a degree of freedom
regarding how the field variables of each element are packed
along a row.  The packing of field variables can be characterised by
considering the distance, $\Delta j$ (in columns) between two subsequent
field variables for a given element.  The case of $\Delta j = 1$
corresponds to the array of structures (AoS) packing whereas the choice
of $\Delta j = \abs{\vec{\Omega}_e}$ leads to the structure of arrays
(SoA) packing.  A hybrid approach wherein $\Delta j = k$ with $k$ being
constant results in the AoSoA($k$) approach.  An implementation is free
to chose between any of these counting patterns so long as it is
consistent. For simplicity PyFR uses the SoA packing order across all platforms.

\subsection{Matrix Multiplication Kernels}

PyFR defers matrix multiplication to the GEMM family of sub-routies
provided a suitable Basic Linear Algebra Subprograms (BLAS)
library. BLAS is available for virtually all platforms and optimised
versions are often maintained by the hardware vendors themselves
(\textit{e.g.} cuBLAS for Nvidia GPUs). This approach greatly
facilitates development of efficient and platform portable code. We
note, however, that the matrix sizes encountered in PyFR are not
necessarily optimal from a GEMM perspective. Specifically, GEMM is
optimised for the multiplication of large square matrices, whereas the
constant operator matrixes in PyFR are `small and square' with
$10$--$100$ rows/columns, and the state matrices are `short and fat'
with $10$--$100$ rows and $10\,000$--$100\,000$ columns. Moreover, we
note that the constant operator matrices are know \textit{a priori}, and
do not change in time. This \textit{a priori} knowledge could, in theory,
be leveraged to design bespoke matrix multiply kernels that are more
efficient than GEMM. Development of such bespoke kernels will be a topic
of future research - with results easily integrated into PyFR as an
optional replacement for GEMM.

\subsection{Point-Wise Kernels}

Point-wise kernels are specified using a domain specific language
implemented in PyFR atop of the Mako templating engine
\cite{mako2013}. The templated kernels are then interpreted at runtime,
converted to low-level code, compiled, linked and loaded. Currently the
templating engine can generate C/OpenMP to target CPUs, and CUDA (via
the PyCUDA wrapper \cite{klockner2012pycuda}) to target Nvidia GPUs. Use
of a domain specific language avoids implementation of each point-wise
kernel for each target platform; keeping the codebase compact and
platform portable. Runtime code generation also means it is possible to
instruct the compiler to emit binaries which are optimised for the
current hardware architecture.  Such optimisations can result in
anything up to a fourfold improvement in performance when compared with
architectural defaults.

\begin{figure}
  \centering
  \includegraphics[scale=0.9]{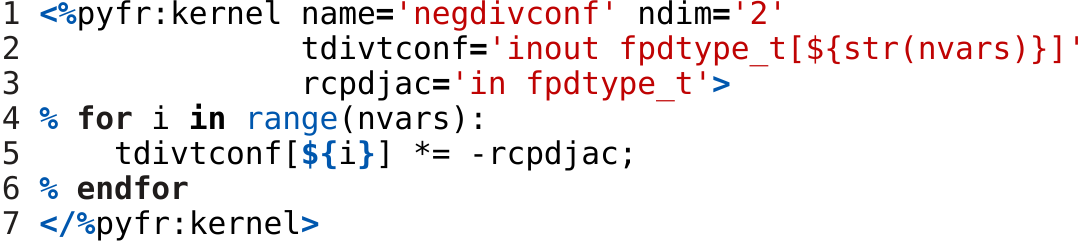}
  \caption{\label{fig:extrinsic-sample}An example of an extrinsic kernel
    in PyFR.  The template variable \emph{nvars} is taken to be the
    number of field variables, $N_v$.  The kernel arguments
    \emph{tdivtconf} and \emph{rcpdjac} correspond to $\tilde{\grad}
    \cdot \tilde{\vec{f}}$ and $J^{-1}$ respectively with the operation
    being performed in-place.}
\end{figure}

\begin{figure}
  \centering
  \includegraphics[scale=0.9]{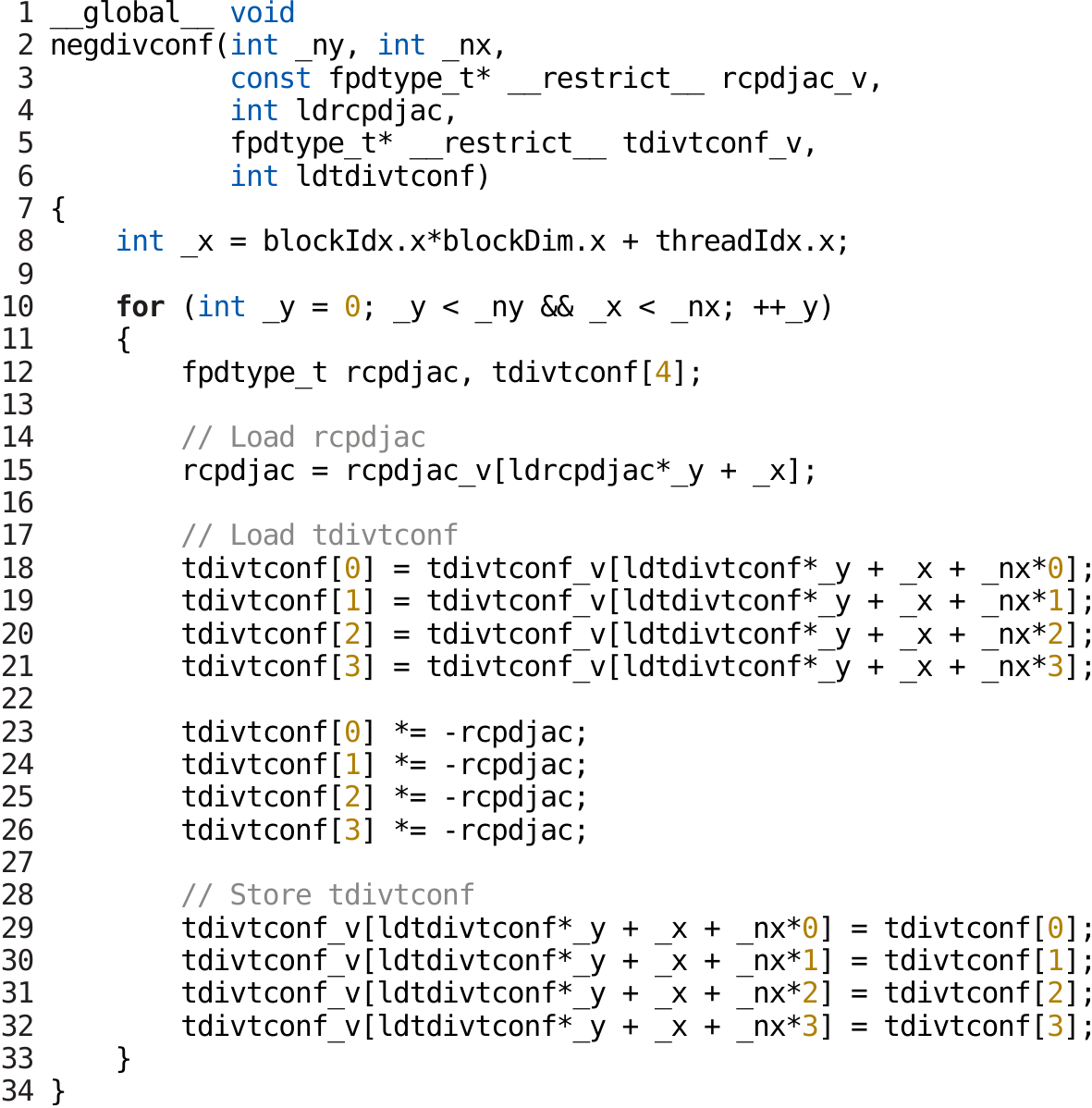}%
  \caption{\label{fig:extrinsic-sample-cuda}Generated CUDA source for the
    template in \autoref{fig:extrinsic-sample} for when $N_V = 4$.}
\end{figure}

\begin{figure}
  \centering
  \includegraphics[scale=0.9]{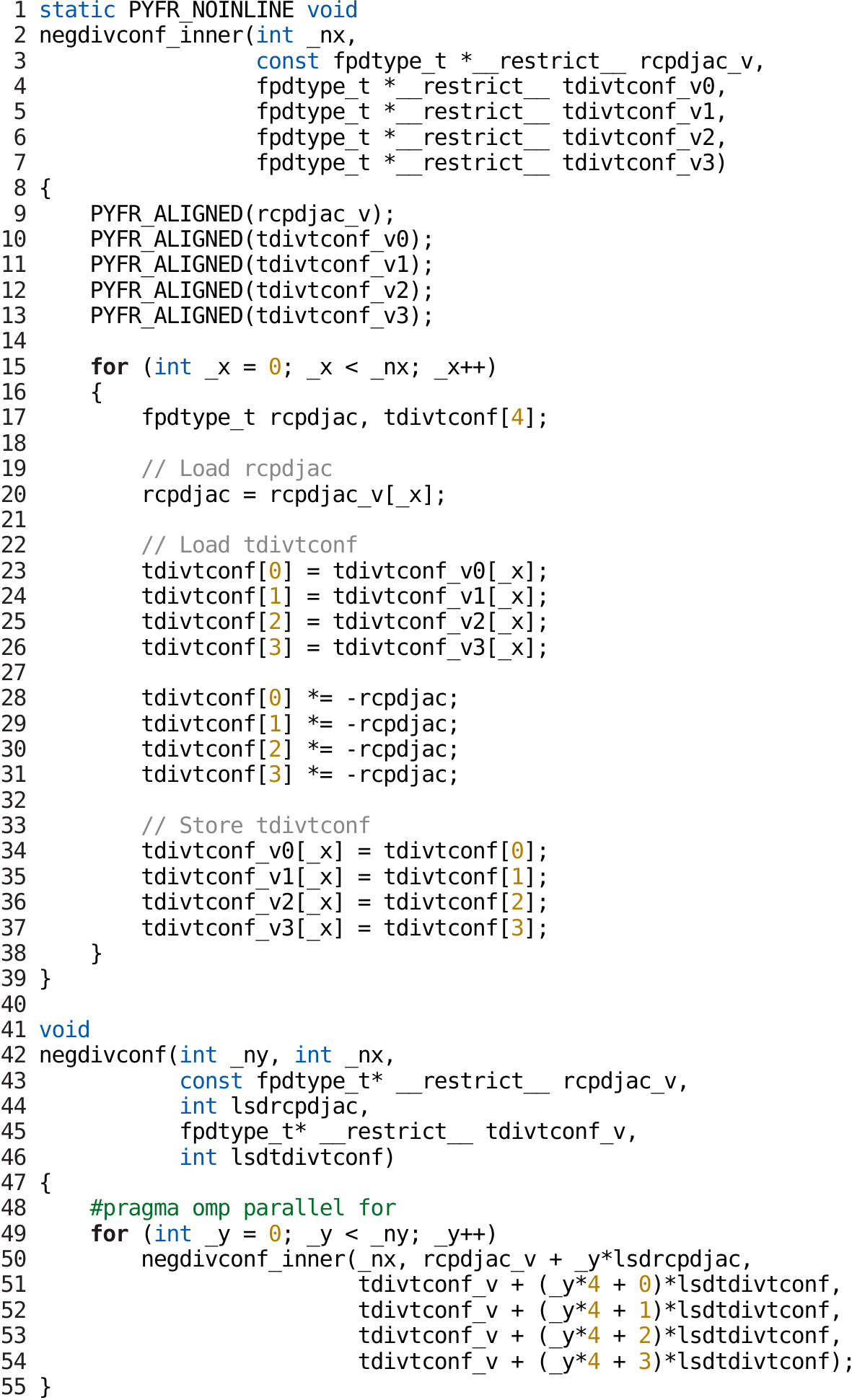}%
  \caption{\label{fig:extrinsic-sample-openmp}Generated OpenMP annotated C
    source code for the template in \autoref{fig:extrinsic-sample} for
    when $N_V = 4$.  The somewhat unconventional structure is necessary
    to ensure that the kernel is properly vectorised across a range of
    compilers.}
\end{figure}

As an example of a point-wise kernel we consider the evaluation of the
right hand side of \autoref{eq:negdivconf}, which reads
$-J^{-1\,(u)}_{e\rho n}(\tilde{\grad} \cdot
\tilde{\vec{f}})^{(u)}_{e\rho n\alpha}$.  The operation consists of a
point-wise multiplication between the negative reciprocal of the
Jacobian and the transformed divergence of the flux at each solution
point.  \autoref{fig:extrinsic-sample} shows how such a kernel can be
expressed in the domain specific language of PyFR. There are several
points of note.  Firstly, the kernel is purely scalar in nature. This is
by design; in PyFR point-wise kernels need only prescribe the point-wise
operation to be applied.  Important choices such as how to vectorise a
given operation or how to gather data from memory are all delegated to
templating engine.  Secondly, we note it is possible to utilise Python
when generating the main body of kernels.  This capability is showcased
on lines four, five and six where it is used to unroll a for loop over
each of the field variables.  Finally, we also highlight the use of an
abstract data type \emph{fpdtype\_t} for floating point variables which
permits a single set of kernels to be used for both single and double
precision operation.  Generated CUDA source for this kernel can be seen
in \autoref{fig:extrinsic-sample-cuda}, and the equivalent C kernel can
be found in \autoref{fig:extrinsic-sample-openmp}.

\subsection{Distributed Memory Parallelism and Scheduling}

PyFR is capable of operating on heigh performance computing clusters utilising distributed
memory parallelism.  This is accomplished through the Message Passing
Interface (MPI).  All MPI functionality is implemented at the Python
level through the mpi4py \cite{mpi4py2013} wrapper.  To enhance the
scalability of the code care has been taken to ensure that all requests
are persistent, point-to-point and non-blocking.  Further, the format of
data that is shared between ranks has been made backend independent.  It
is therefore possible to deploy PyFR on heterogeneous clusters
consisting of both conventional CPUs and accelerators.

\begin{figure}
  \centering
  \includegraphics[width=8cm]{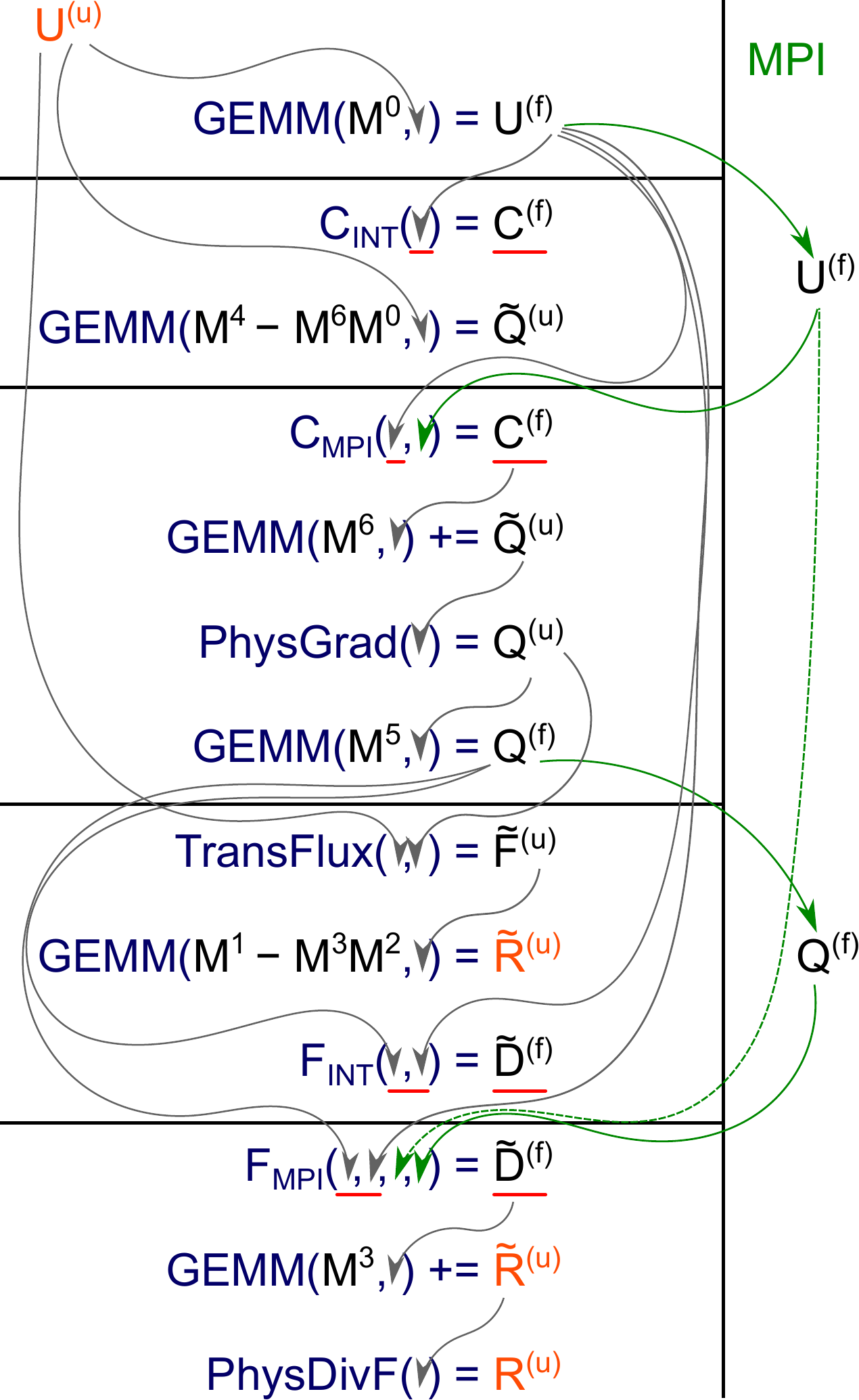}
  \caption{\label{fig:flow}Flow diagram showing the stages required to
    compute $-\grad \cdot \vec{f}$.  Symbols correspond to those of
    \autoref{sec:mat-rep}.  For simplicity arguments referencing
    constant data have been omitted.  Memory indirection is indicated by
    red underlines.  Synchronisation points are signified by black
    horizontal lines.  Dotted lines correspond to data reuse.}
\end{figure}

The arrangement of kernel calls required to solve an advection-diffusion problem can be seen in \autoref{fig:flow}. Our primary objective when scheduling kernels was to maximise the potential for
overlapping communication with computation. In order to help achieve this the common
interface solution, $\mathfrak{C}_{\alpha}$, and common interface flux,
$\mathfrak{F}_{\alpha}$, kernels have been broken apart into two
separate kernels; suffixed in the figure by \textsc{int} and
\textsc{mpi}. PyFR is therefore able to perform a significant degree of
rank-local computation while the relevant ghost states are being
exchanged.

Our secondary objective when scheduling kernels was to minimise the
amount of temporary storage required during the evaluation of $-\grad
\cdot \vec{f}$.  Such optimisations are critical within the context of
accelerators which often have an order of magnitude less memory than a
contemporary platform.  In order to help achieve this $\mat{U}^{(u)}$,
$\tilde{\mat{R}}^{(u)}$, and $\mat{R}^{(u)}$ are allowed to alias.  By
permitting the same storage location to be used for both the inputted
solution and the outputted flux divergence it is possible to reduce the
storage requirements of the RK schemes.  Another opportunity for memory
reuse is in the transformed flux function where the incoming gradients,
$\mat{Q}^{(u)}$, can be overwritten with the transformed flux,
$\tilde{\mat{F}}^{(u)}$.  A similar approach can be used in the common
interface flux function whereby $\mat{U}^{(f)}$ can updated in-place
with the entires of $\tilde{\mat{D}}^{(f)}$ which holds the transformed
common normal flux.  Moreover, $\mat{C}^{(f)}$ is also able to utilise
the same storage as the somewhat larger $\mat{Q}^{(f)}$ array.  These
optimisations allow PyFR to process over $100\,000$ curved,
unstructured, hexahedral elements at $\wp = 3$ inside of a
$5\,\text{GiB}$ memory footprint.

\section{Governing Systems}
\label{sec:govsystems}

\subsection{Overview}

PyFR is a framework for solving various advection-diffusion type problems. In the current release of PyFR two specific governing systems can be solved, specifically the Euler equations for inviscid compressible flow, and the compressible Navier-Stokes equations for viscous compressible flow. Details regarding both are given below.

\subsection{Euler Equations}

Using the framework introduced in \autoref{sec:discretisation} the three
dimensional Euler equations can be expressed in conservative form as
\begin{equation} \label{eq:eulerflux}
  u = \begin{Bmatrix}
    \rho\\
    \rho v_x\\
    \rho v_y\\
    \rho v_z\\
    E
  \end{Bmatrix}, \qquad
  \vec{f}  = \vec{f}^{(\mathrm{inv})} = \begin{Bmatrix}
    \rho v_x       & \rho v_y       & \rho v_z\\
    \rho v_x^2 + p & \rho v_yv_x    & \rho v_zv_x\\
    \rho v_xv_y    & \rho v_y^2 + p & \rho v_zv_y\\
    \rho v_xv_z    & \rho v_yv_z    & \rho v_z^2 + p\\
    v_x(E + p)     & v_y(E + p)     & v_z(E + p)
  \end{Bmatrix},
\end{equation}
with $u$ and $\vec{f}$ together satisfying \autoref{eq:advdiffpde}.  In
the above $\rho$ is the mass density of the fluid, $\vec{v}
=(v_x,v_y,v_z)^T$ is the fluid velocity vector, $E$ is the total energy
per unit volume and $p$ is the pressure.  For a perfect gas the pressure
and total energy can be related by the ideal gas law
\begin{equation} \label{eq:idealgas}
  E = \frac{p}{\gamma - 1} + \frac{1}{2}\rho\|\vec{v}\|^2,
\end{equation}
with $\gamma = C_p/C_v$.

With the fluxes specified all that remains is to prescribe a method for
computing the common normal flux, $\mathfrak{F}_{\alpha}$, at interfaces
as defined in \autoref{eq:commpflux}.  This can be accomplished using an
approximate Riemann solver for the Euler equations.  There exist a
variety of such solvers as detailed in \cite{toro2009riemann}.  A
description of those implemented in PyFR can be found in
\autoref{sec:riemann-solvers}.

\subsection{Compressible Navier-Stokes Equations}

The compressible Navier-Stokes equations can be viewed as an extension of the Euler
equations via the inclusion of viscous terms.  Within the framework outlined above the
flux now takes the form of $\vec{f} = \vec{f}^{(\text{inv})} -
\vec{f}^{(\text{vis})}$ where
\begin{equation} \label{eq:visflux} \vec{f}^{(\mathrm{vis})}
  = \begin{Bmatrix}
    0 & 0 & 0\\
    \mathcal{T}_{xx} & \mathcal{T}_{yx} & \mathcal{T}_{zx}\\
    \mathcal{T}_{xy} & \mathcal{T}_{yy} & \mathcal{T}_{zy}\\
    \mathcal{T}_{xz} & \mathcal{T}_{yz} & \mathcal{T}_{zz}\\
    v_i\mathcal{T}_{ix} + \Delta\partial_x T & v_i\mathcal{T}_{iy} +
    \Delta\partial_y T & v_i\mathcal{T}_{iz} + \Delta\partial_z T
    \end{Bmatrix}.
\end{equation}
In the above we have defined $\Delta = \mu C_p/P_r$ where $\mu$ is the
dynamic viscosity and $P_r$ is the Prandtl number.  The components of
the stress-energy tensor are given by
\begin{equation}
  \mathcal{T}_{ij} = \mu(\partial_i v_j + \partial_j v_i) -
  \frac{2}{3}\mu\delta_{ij}\grad \cdot \vec{v}.
\end{equation}
Using the ideal gas law the temperature
can be expressed as
\begin{equation}
T = \frac{1}{C_v}\frac{1}{\gamma -1}\frac{p}{\rho},
\end{equation}
with partial derivatives thereof being given according to the quotient
rule.

Since the Navier-Stokes equations are an advection-diffusion type system
it is necessary to both compute a common solution
($\mathfrak{C}_{\alpha}$ of \autoref{eq:commsoln}) at element boundaries
and augment the inviscid Riemann solver to handle the viscous part of
the flux. A popular approach is the LDG method as presented in
\cite{hesthaven2008nodal,castonguay2013energy}.  In this approach the
common solution is given $\forall \alpha$ according to
\begin{equation}
  \mathfrak{C}(u_L,u_R) = (\tfrac{1}{2} - \beta)u_L + (\tfrac{1}{2} + \beta)u_R,
\end{equation}
where $\beta$ controls the degree of upwinding/downwinding.  The common
normal interface flux is then prescribed, once again $\forall \alpha$,
according to
\begin{equation}
  \mathfrak{F}(u_L,u_R,\vec{q}_L, \vec{q}_R, \hat{\vec{n}}_L) =
  \mathfrak{F}^{(\text{inv})} - \mathfrak{F}^{(\text{vis})},
\end{equation}
where $\mathfrak{F}^{(\text{inv})}$ is a suitable inviscid Riemann
solver (see \autoref{sec:riemann-solvers}) and
\begin{equation}
  \mathfrak{F}^{(\text{vis})} = \hat{\vec{n}}^{\vphantom{(\text{vis})}}_L \cdot \left\{(\tfrac{1}{2} + \beta)\vec{f}^{(\text{vis})}_L +
    (\tfrac{1}{2} - \beta)\vec{f}^{(\text{vis})}_R\right\}
  + \tau(u_L^{\vphantom{(\text{vis})}} - u_R^{\vphantom{(\text{vis})}}),
\end{equation}
with $\tau$ being a penalty parameter, $\vec{f}^{(\text{vis})}_L =
\vec{f}^{(\text{vis})}_{\vphantom{L}}(u^{\vphantom{(\text{vis})}}_L,\vec{q}^{\vphantom{(\text{vis})}}_L)$,
and $\vec{f}^{(\text{vis})}_R =
\vec{f}^{(\text{vis})}_{\vphantom{R}}(u^{\vphantom{(\text{vis})}}_R,\vec{q}^{\vphantom{(\text{vis})}}_R)$.
We observe here that if the common solution is upwinded then the common
normal flux will be downwinded.  Generally, $\beta = \pm1/2$ as
this results in the numerical scheme having a compact stencil and $0 \le
\tau \le 1$.

\subsubsection{Presentation in Two Dimensions}

The above prescriptions of the Euler and Navier-Stokes equations are
valid for the case of $N_D = 3$.  The two dimensional formulation can be
recovered by deleting the fourth rows in the definitions of $u$,
$\vec{f}^{(\text{inv})}$ and $\vec{f}^{(\text{vis})}$ along with the
third columns of $\vec{f}^{(\text{inv})}$ and $\vec{f}^{(\text{vis})}$.
Vectors are now two dimensional with the velocity being given by
$\vec{v} = (v_x,v_y)^T$.

\section{Validation}
\label{sec:validation}

\subsection{Euler Equations: Euler Vortex Super Accuracy}

Various authors \cite{huynh2007flux,vincent2011insights} have shown FR
schemes exhibit so-called `super accuracy' (an order of accuracy greater
than the expected $\wp + 1$).  To confirm PyFR can achieve super
accuracy for the Euler equations a square domain $\vec{\Omega} =
[-20,20]^2$ was decomposed into four structured quad meshes with
spacings of $h =1/3$, $h = 2/7$, $h = 1/4$, and $h =2/9$.  Initial
conditions were taken to be those of an isentropic Euler vortex in a
free-stream
\begin{align}
  \rho(\vec{x},t=0) &= \left\{1 - \frac{S^2M^2(\gamma -
      1)\exp 2f}{8\pi^2}\right\}^{\frac{1}{\gamma -1}},\\
  \vec{v}(\vec{x},t=0) &= \frac{Sy\exp{f}}{2\pi R}\hat{\vec{x}} + \left\{1 - \frac{Sx\exp{f}}{2\pi R}\right\}\hat{\vec{y}},\\
  p(\vec{x},t=0) &= \frac{\rho^{\gamma}}{\gamma M^2},
\end{align}
where $f = (1 - x^2 - y^2)/2R^2$, $S = 13.5$ is the strength of the
vortex, $M = 0.4$ is the free-stream Mach number, and $R = 1.5$ is the
radius.  All meshes were configured with periodic boundary conditions
along boundaries of constant $x$.  Along boundaries of constant $y$ the
dynamical variables were fixed according to
\begin{align*}
  \rho(\vec{x} = x\hat{\vec{x}} \pm20\hat{\vec{y}},t) &= 1,\\
  \vec{v}(\vec{x} = x\hat{\vec{x}} \pm20\hat{\vec{y}},t) &= \hat{\vec{y}},\\
  p(\vec{x} = x\hat{\vec{x}} \pm20\hat{\vec{y}},t) &= \frac{1}{\gamma M^2},
\end{align*}
which are simply the limiting values of the initial conditions.
Strictly speaking these conditions, on account of the periodicity,
result in the modelling of an infinite array of coupled vortices.  The
impact of this is mitigated by the observation that the exponentially
decaying vortex has a characteristic radius which is far smaller than
the extent of the domain.  Neglecting these effects the analytic solution
of the system is a time $t$ is simply a translation of the initial
conditions.

\begin{figure}
  \centering
  \includegraphics{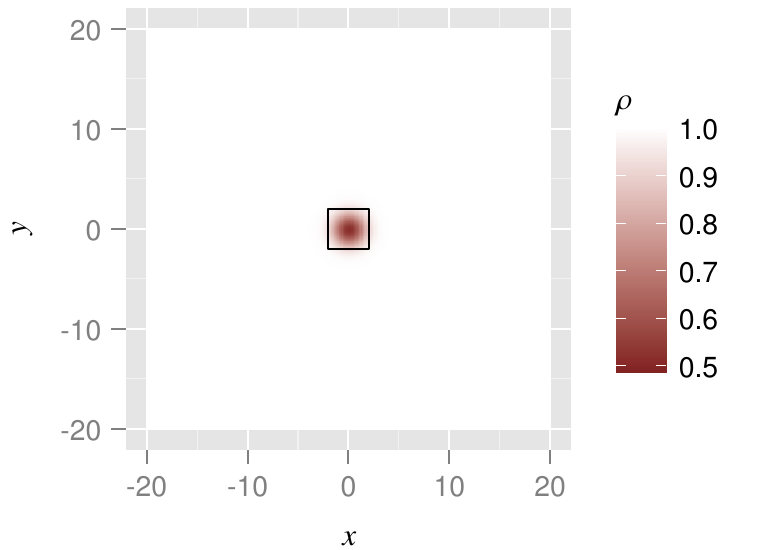}%
  \caption{\label{fig:vortex-ic}Initial density profile for the vortex
    in $\vec{\Omega}$.  The black box shows the region where the error
    is calculated.}
\end{figure}

Using the analytical solution we can define an $L^2$ error as
\begin{equation} \label{eq:sigma}
  \sigma(t)^2 = \int_{-2}^{2} \int_{-2}^{2}\Bigl[\rho^{\delta}(\vec{x} +
  \Delta_y(t)\hat{\vec{y}},t) - \rho(\vec{x}, t = 0)\Bigr]^2 \udds \vec{x},
\end{equation}
where $\rho^{\delta}(\vec{x},t)$ is the numerical mass density,
$\rho(\vec{x},t=0)$ the analytic mass density, and $\Delta_y(t)$ is the
ordinate corresponding to the centre of the vortex at a time $t$ and
accounts for the fact that the vortex is translating in a free stream
velocity of unity in the $y$ direction.  Restricting the region of
consideration to a small box centred around the origin serves to further
mitigate against the effects of vortices coupling together.  The initial
mass density along with the $[-2,-2] \times [2,2]$ region used to
evaluate the error can be seen in \autoref{fig:vortex-ic}.  At times,
$t_c$, when the vortex is centred on the box the error can be readily
computed by integrating over each element inside the box and summing the
residuals
\begin{equation}
  \sigma(t_c)^2 =
  \iint_{\hat{\vec{\Omega}}_e} \Bigl[\rho^{\delta}_{i}(\tilde{\vec{x}}, t_c) -
  \rho(\bm{\mathcal{M}}_{i}(\tilde{\vec{x}}), 0)\Bigr]^2 J_i(\tilde{\vec{x}})
  \udds\tilde{\vec{x}},
\end{equation}
where, $\rho^{\delta}_{i}(\tilde{\vec{x}}, t_c)$ is the approximate mass
density inside of the $i$th element, and $J_i(\tilde{\vec{x}})$ the
associated Jacobian.  These integrals can be approximated by applying
Gaussian quadrature
\begin{equation}
\begin{aligned}
  \sigma(t_c)^2 &\approx
  J_i(\tilde{\vec{x}}_j)\Bigl[\rho^{\delta}_{i}(\tilde{\vec{x}}_j, t_c)
  - \rho(\bm{\mathcal{M}}_{i}(\tilde{\vec{x}}_j), 0)\Bigr]^2
  \omega_j\\
  &= \frac{h^2}{4} \Bigl[\rho^{\delta}_{i}(\tilde{\vec{x}}_j, t_c) -
  \rho(\bm{\mathcal{M}}_{i}(\tilde{\vec{x}}_j), 0)\Bigr]^2 \omega_j,
\end{aligned}
\end{equation}
where $\set{\tilde{\vec{x}}_j}$ are abscissa and $\set{\omega_j}$ the
weights of a rule determined for integration inside of a standard
quadrilateral.  So long as the rule used is of a suitable strength then
this will be a very good approximation of the true $L^2$ error.

\begin{figure}
  \centering
  \includegraphics{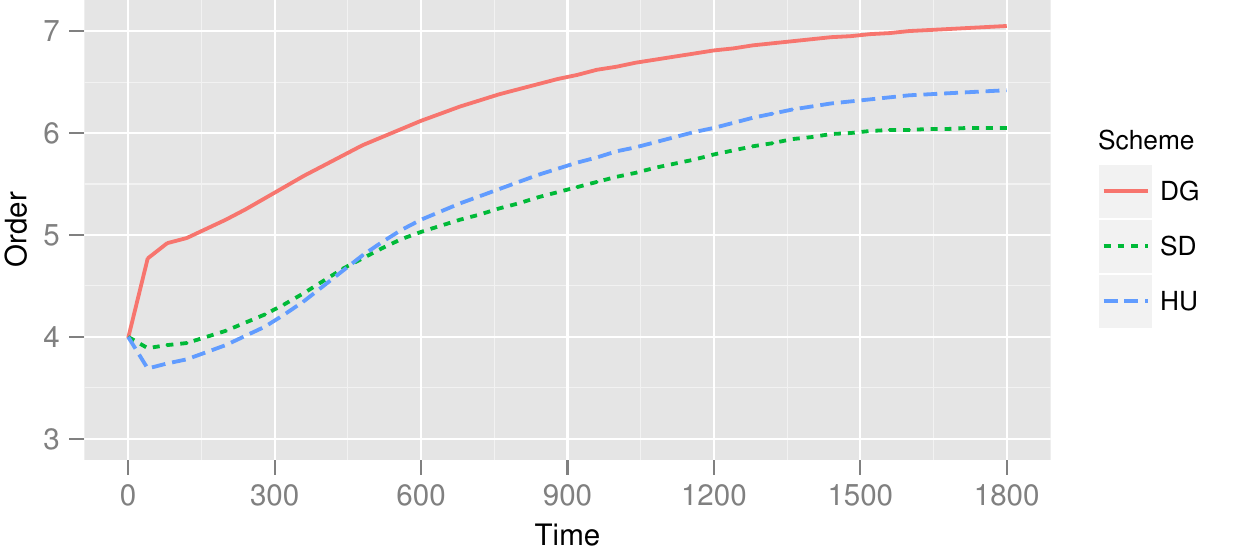}%
  \caption{\label{fig:superacc}Spatial super accuracy observed for a
    $\wp = 3$ simulation using DG, SD and HU as defined in \cite{vincent2011insights}.}
\end{figure}

Following \cite{vincent2011insights} the initial conditions were laid
onto the mesh using a collocation projection with $\wp = 3$.  The
simulation was then run with three different flux reconstruction
schemes: DG, SD, and HU as defined in \cite{vincent2011insights}.
Solution points were placed at a tensor product construction of
Gauss-Legendre quadrature points.  Common interface fluxes were computed
using a Rusanov Riemann solver.  To advance the solutions in time a
classical fourth order Runge-Kutta method (RK4) was used.  The time step
was taken to be $\Delta t = 0.00125$ with $t = 0..1800$ with solutions
written out to disk every $32\,000$ steps.  The order of accuracy of the
scheme at a particular time can be determined by plotting $\log \sigma$
against $\log h$ and performing a least-squares fit through the four
data points.  The order is given by the gradient of the fit.  A plot of
order of accuracy against time for the three schemes can be seen in
\autoref{fig:superacc}.  We note that the order of accuracy changes as a
function of time.  This is due to the fact that the error is actually of
the form $\sigma(t) = \sigma_{\text{p}} + \sigma_{\text{so}}(t)$ where
$\sigma_{\text{p}}$ is a constant projection error and
$\sigma_{\text{so}}$ is a time-dependent spatial operator error. The
projection error arises as a consequence of the forth order collocation
projection of the initial conditions onto the mesh.  Over time the
spatial operator error grows in magnitude and eventually dominates.
Only when $\sigma_{\text{so}}(t) \gg \sigma_{\text{p}}$ can the true
order of the method be observed.  The results here can be seen to be in
excellent agreement with those of \cite{vincent2011insights}.

\subsection{Compressible Navier-Stokes Equations: Couette Flow}

\begin{figure}
  \centering
  \includegraphics{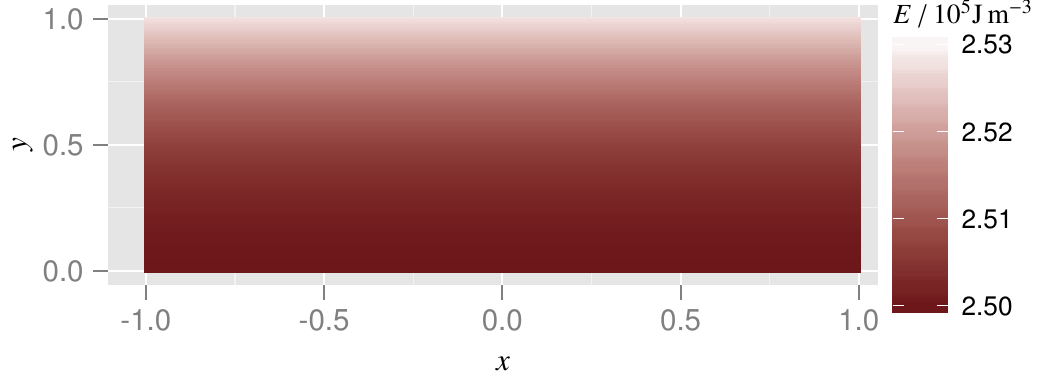}%
  \caption{\label{fig:couette-energy}Converged steady state energy
    profile for the two dimensional Couette flow problem.}
\end{figure}

Consider the case in which two parallel plates of infinite extent are
separated by a distance $H$ in the $y$ direction.  We treat both plates
as isothermal walls at a temperature $T_w$ and permit the top plate to
move at a velocity $v_w$ in the $x$ direction with respect to the bottom
plate.  For simplicity we shall take the ordinate of the bottom plate as
zero.  In the case of a constant viscosity $\mu$ the Navier-Stokes
equations admit an analytical solution in which

\begin{align}
  \rho(\phi) &= \frac{\gamma}{\gamma - 1} \frac{2p}{2C_pT_w + P_r v_w^2\phi
    (1 - \phi)},\\
  \vec{v}(\phi) &= v_w\phi\hat{\vec{x}},\\
  p &= p_c,
\end{align}
where $\phi = y/H$ and $p_c$ is a constant pressure.  The total energy
is given by the ideal gas law of \autoref{eq:idealgas}.  On a finite
domain the Couette flow problem can be modelled through the imposition
of periodic boundary conditions.  For a two dimensional mesh periodicity
is enforced in $x$ whereas for three dimensional meshes it is enforced
in both $x$ and $z$.  To validate the Navier-Stokes solver in PyFR we
take $\gamma = 1.4$, $P_r = 0.72$, $\mu = 0.417$, $C_p =
\SI{1005}{\joule\per\kelvin}$, $H = \SI{1}{\metre}$, $T_w =
\SI{300}{\kelvin}$, $p_c = \SI{1e5}{\pascal}$, and $v_w =
\SI{69.445}{\metre\per\second}$.  These values correspond to a Mach
number of 0.2 and a Reynolds number of 200.  The plates were modelled as
no-slip isothermal walls as detailed in
\autoref{sec:bc-no-slip-isot-wall} of \autoref{sec:boundary-conditions}.
A plot of the resulting energy profile can be seen in
\autoref{fig:couette-energy}.  Constant initial conditions are taken as
$\rho = \big\langle\,\rho(\phi)\,\big\rangle$, $\vec{v} =
v_w\hat{\vec{x}}$, and $p = p_c$.  Using the analytical solution we
again define an $L^2$ error as
\begin{align}
  \sigma(t)^2 &= \int_{\vec{\Omega}} \left[E^{\delta}(\vec{x},t) -
    E(\vec{x})\right]^2 \uds^{N_D} \vec{x}\\
  &= \int_{\vec{\Omega}_{ei}} \left[E^{\delta}_{ei}(\tilde{\vec{x}},t) -
    E(\bm{\mathcal{M}}_{ei}(\tilde{\vec{x}}))\right]^2
  J_{ei}(\tilde{\vec{x}}) \uds^{N_D} \tilde{\vec{x}}\\
  &\approx \left[E^{\delta}_{ei}(\tilde{\vec{x}}_{ej},t) -
    E(\bm{\mathcal{M}}_{ei}(\tilde{\vec{x}}_{ej}))\right]^2
  J_{ei}(\tilde{\vec{x}}_{ej})\omega_{ej},
\end{align}
where $\vec{\Omega}$ is the computational domain,
$E^{\delta}(\vec{x},t)$ is the numerical total energy, and $E(\vec{x})$
the analytic total energy.  In the third step we have approximated each
integral by a quadrature rule with abscissa $\set{\tilde{\vec{x}}_{ej}}$
and weights $\set{\omega_{ej}}$ inside of an element type $e$.  Couette
flow is a steady state problem and so in the limit of $t \rightarrow
\infty$ the numerical total energy should converge to a solution.
Starting from a constant initial condition the $L^2$ error was computed
every $0.1$ time units.  The simulation was said to have converged when
$\sigma(t)/\sigma(t + 0.1) \le 1.01$ where $\sigma$ is the $L^2$ error.
We will denote the time at which this occurs by $t_{\infty}$.

Once the system has converged for a range of meshes it is possible to
compute the order of accuracy of the scheme.  For a given $\wp$ this is
the slope (plus or minus a standard error) of a linear least squares fit
of $\log h \sim \log \sigma(t_\infty)$ where $h$ is an approximation of
the characteristic grid spacing.  The expected order of accuracy is $\wp
+ 1$.  In all simulations inviscid fluxes were computed using the
Rusanov approach and the LDG parameters were taken to be $\beta = 1/2$
and $\tau = 0.1$.  All simulations were performed with DG correction
functions and at double precision.  Inside tensor product elements
Gauss-Legendre solution and flux points were employed.  Triangular
elements utilised Williams-Shunn solution points and Gauss-Legendre flux
points.

\paragraph{Two dimensional unstructured mixed mesh.}
For the two dimensional test cases the computational domain was taken to
be $[-1,1] \times [0,1]$.  This domain was then meshed with both
triangles and quadrilaterals at four different refinement levels.  The
Couette flow problem described above was then solved on each of these
meshes.  Experimental $L^2$ errors and orders of accuracy can be seen in
\autoref{tab:couette-2d-us}.  We note that in all cases the expected
order of accuracy was obtained.

\begin{figure}
  \centering
  \begin{subfigure}[b]{.45\linewidth}
    \centering
    \includegraphics[width=5cm]{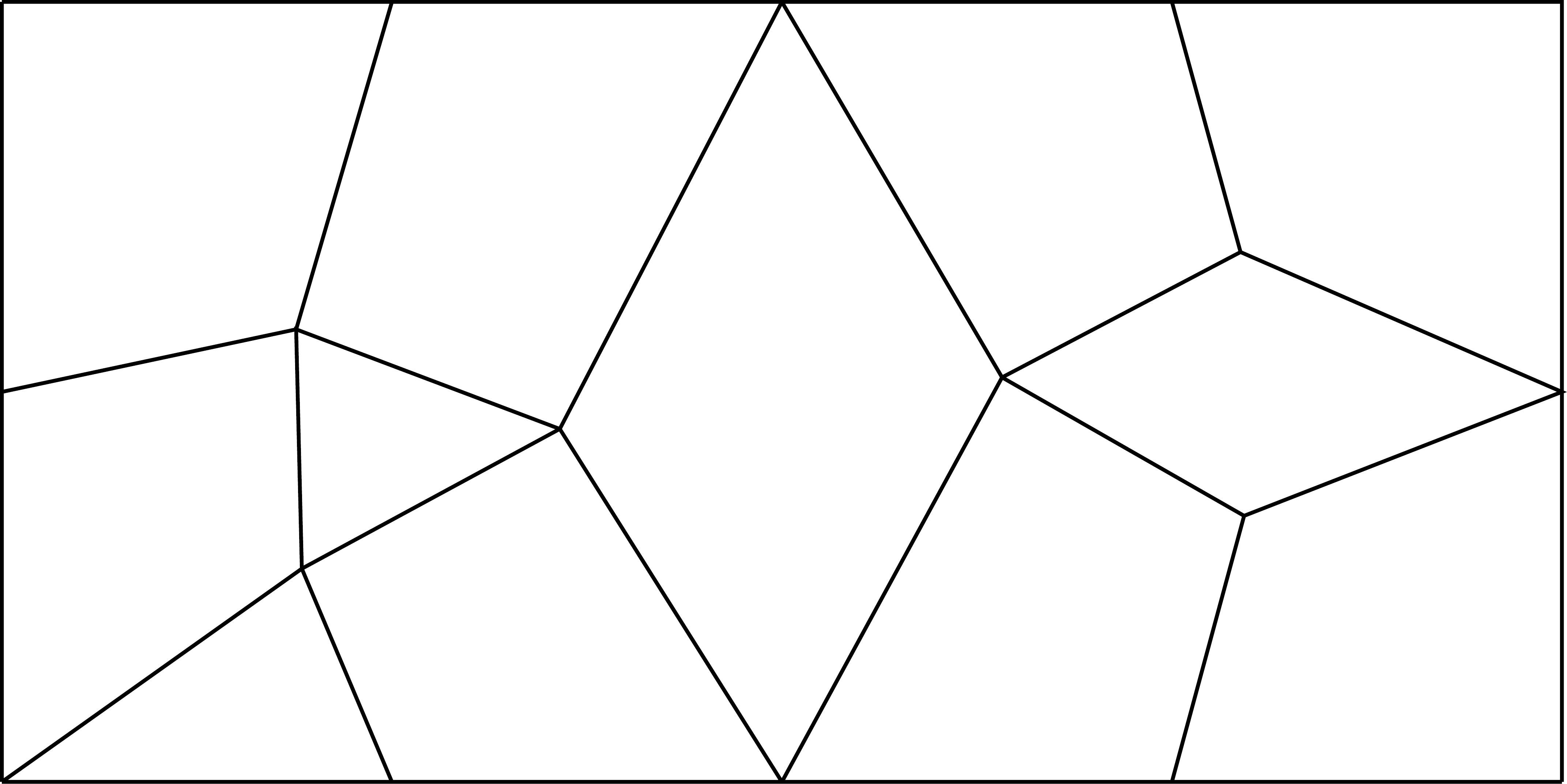}
    \caption{}
  \end{subfigure}
  \begin{subfigure}[b]{.45\linewidth}
    \centering
    \includegraphics[width=5cm]{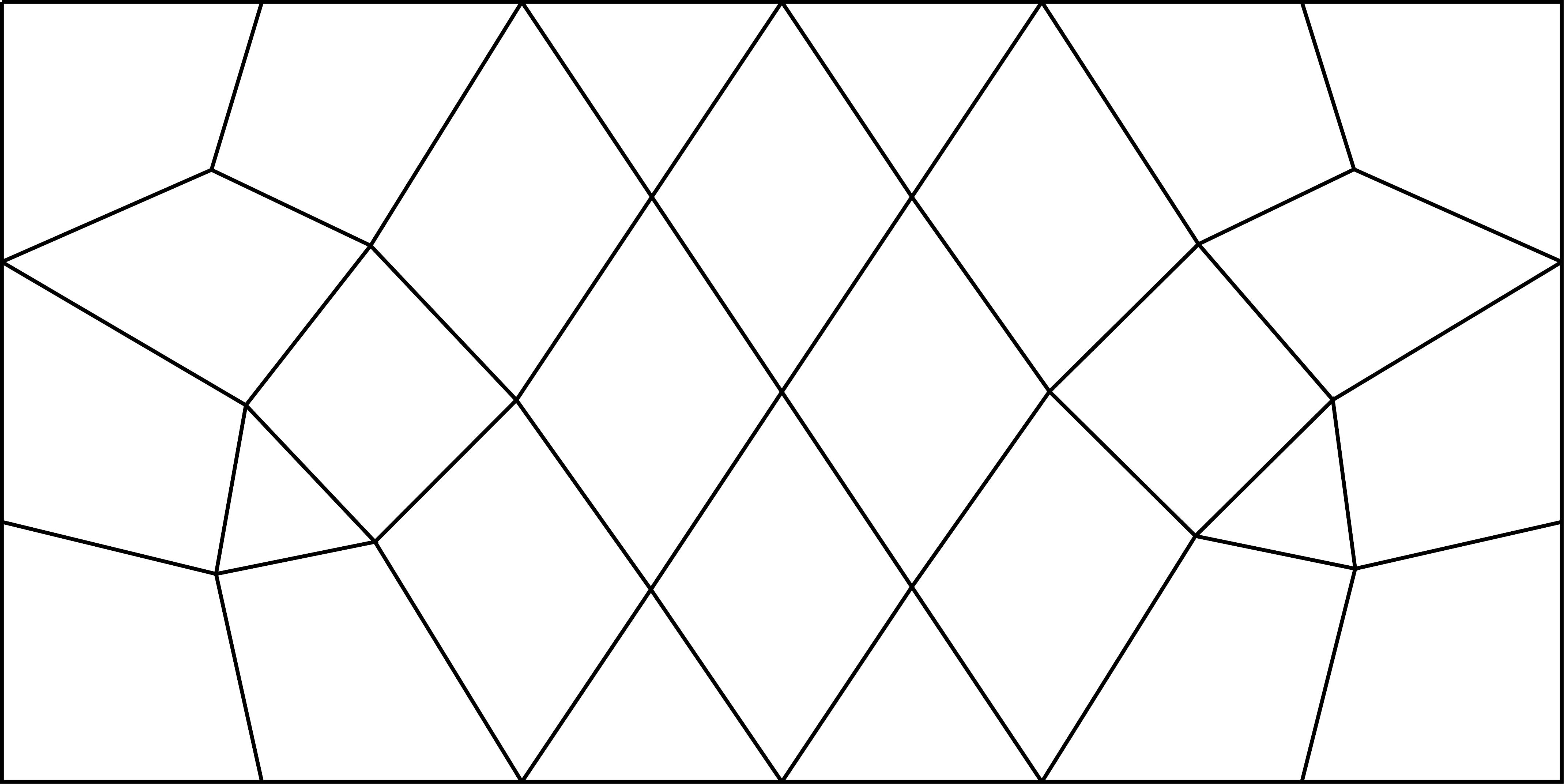}
    \caption{}
  \end{subfigure}\vskip12pt
  \begin{subfigure}[b]{.45\linewidth}
    \centering
    \includegraphics[width=5cm]{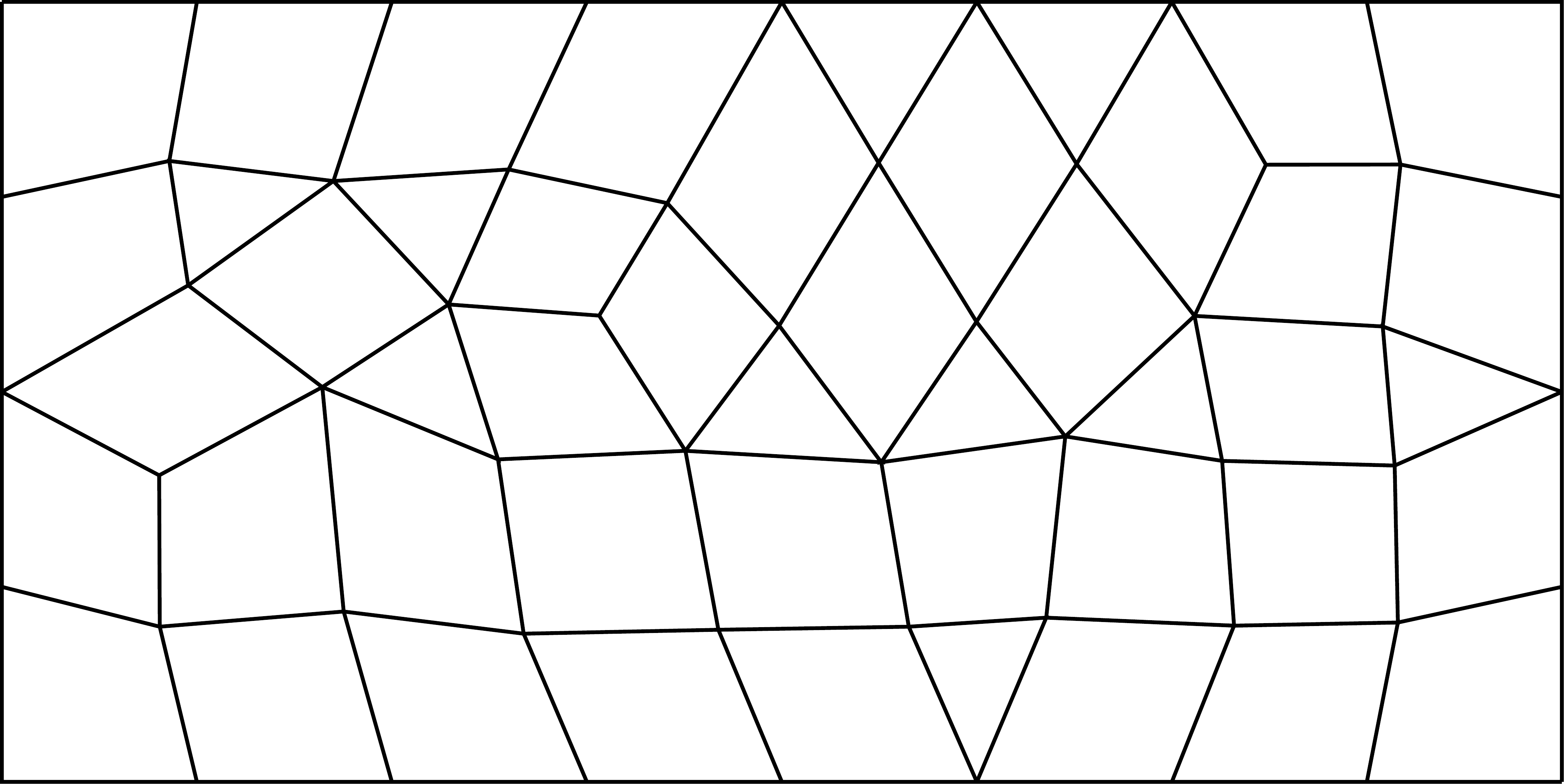}
    \caption{}
  \end{subfigure}
  \begin{subfigure}[b]{.45\linewidth}
    \centering
    \includegraphics[width=5cm]{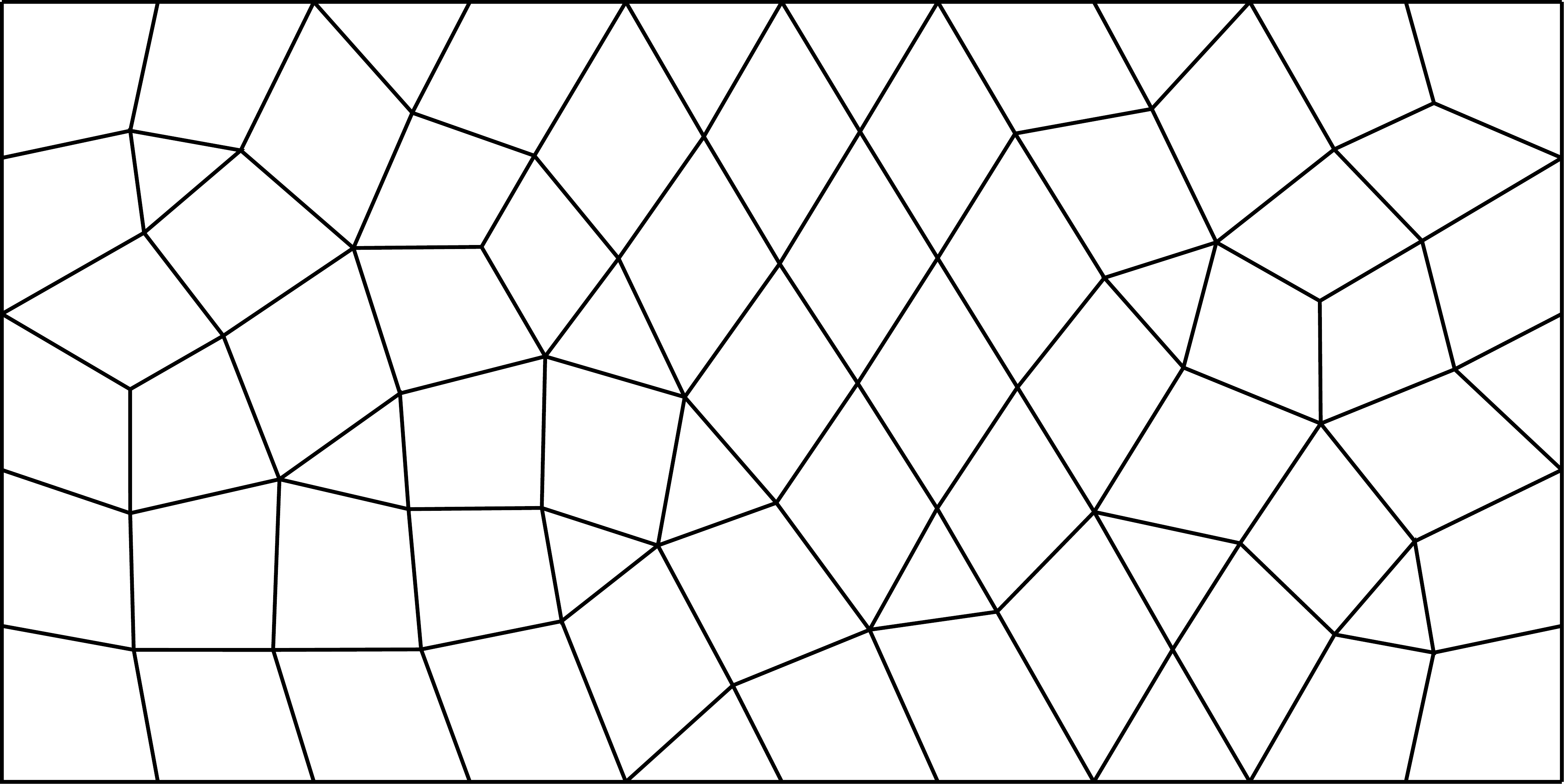}
    \caption{}
  \end{subfigure}
  \caption{Unstructured mixed element meshes used for the two
    dimensional Couette flow problem.}
\end{figure}
\begin{table}
  \centering
  \caption{\label{tab:couette-2d-us}$L^2$ energy error and orders of
    accuracy for the Couette flow problem on four mixed meshes.  The
    mesh spacing  was approximated as $h \sim N_E^{-1/2}$ where $N_E$ is
    the total number of elements in the mesh.}
  \begin{tabular}{rrllll} \toprule
     & & \multicolumn{4}{c}{$\sigma(t_\infty)\,/\,\si{\joule\per\metre\cubed}$} \\
    \cmidrule{3-6}
    \ccol{Tris} & \ccol{Quads} & \ccol{$\wp=1$} & \ccol{$\wp = 2$} &
    \ccol{$\wp = 3$} & \ccol{$\wp = 4$} \\
    \midrule
     2 &  8 & $1.26 \times 10^{2}$ & $5.77 \times 10^{-1}$ & $5.54 \times 10^{-3}$ & $6.62 \times 10^{-5}$ \\
     6 & 22 & $3.56 \times 10^{1}$ & $1.40 \times 10^{-1}$ & $6.72 \times 10^{-4}$ & $3.91 \times 10^{-6}$ \\
    10 & 37 & $2.08 \times 10^{1}$ & $4.35 \times 10^{-2}$ & $2.54 \times 10^{-4}$ & $8.16 \times 10^{-7}$ \\
    16 & 56 & $1.46 \times 10^{1}$ & $3.52 \times 10^{-2}$ & $1.09 \times 10^{-4}$ & $4.62 \times 10^{-7}$ \\ \midrule
    \multicolumn{2}{r}{Order}
            & {$2.21 \pm 0.12$} & {$2.99 \pm 0.32$}
            & {$3.97 \pm 0.05$} & {$5.20 \pm 0.38$} \\ \bottomrule
  \end{tabular}
\end{table}

\paragraph{Three dimensional extruded hexahedral mesh.}
For this three dimensional case the computational domain was taken to be
$[-1,1] \times [0,1] \times [0,1]$.  Meshes were constructed through
first generating a series of unstructured quadrilateral meshes in the
$x$-$y$ plane.  A three layer extrusion was then performed on this
meshes to yield a series of hexahedral meshes.  Experimental $L^2$
errors and orders of accuracy for these meshes can be seen in
\autoref{tab:couette-3d-ex}.

\begin{table}
  \centering
  \caption{\label{tab:couette-3d-ex}$L^2$ energy errors and orders of
    accuracy for the Couette flow problem on three extruded hexahedral
    meshes. On account of the extrusion $h \sim N^{-1/2}_E$ where $N_E$
    is the total number of elements in the mesh.}
  \begin{tabular}{rlll} \toprule
    & \multicolumn{3}{c}{$\sigma(t_\infty)\,/\,\si{\joule\per\metre\cubed}$} \\
    \cmidrule{2-4}
    \ccol{Hexes} & \ccol{$\wp=1$} & \ccol{$\wp = 2$} &
    \ccol{$\wp = 3$} \\
    \midrule
      78 & $3.35 \times 10^{1}$ & $5.91 \times 10^{-2}$ & $7.28 \times 10^{-4}$ \\
     195 & $1.23 \times 10^{1}$ & $1.87 \times 10^{-2}$ & $1.15 \times 10^{-4}$ \\
     405 & $6.15 \times 10^{0}$ & $5.49 \times 10^{-3}$ & $2.72 \times 10^{-5}$ \\ \midrule
    \multicolumn{1}{r}{Order}
         & {$2.06 \pm 0.08$} & {$2.87 \pm 0.24$}
         & {$3.99 \pm 0.03$} \\ \bottomrule
  \end{tabular}
\end{table}

\begin{figure}
  \centering
  \includegraphics[width=6.5cm]{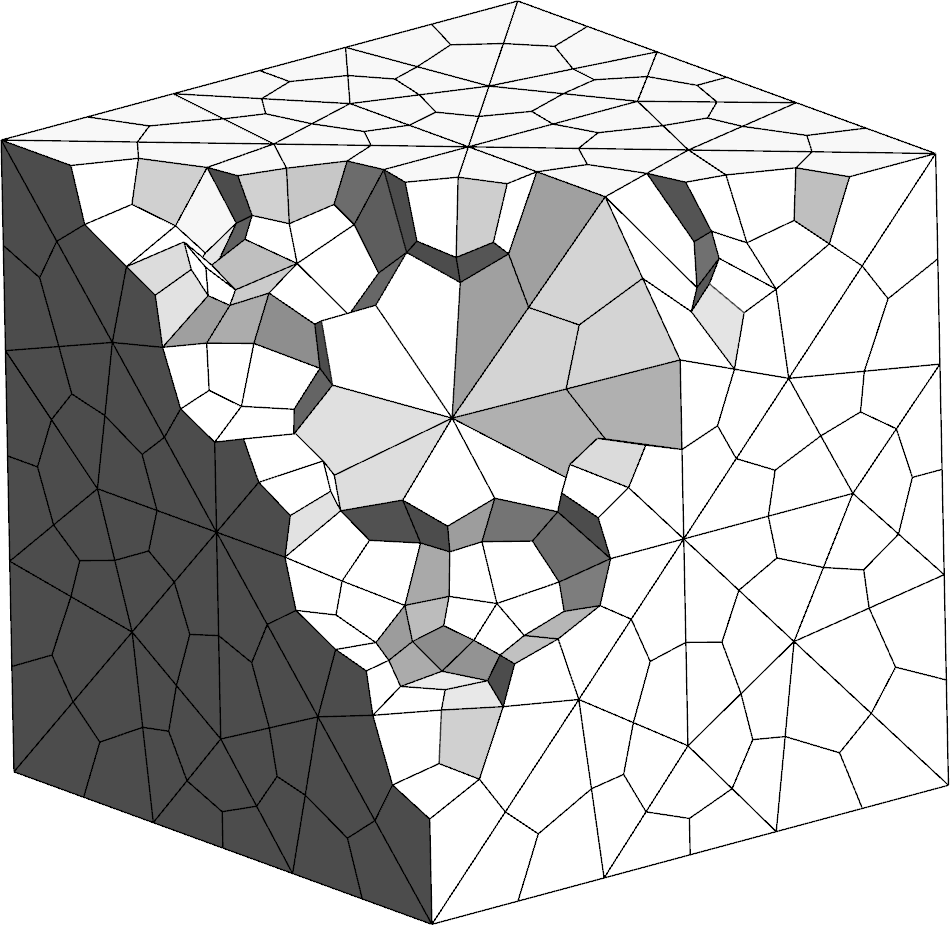}
  \caption{\label{fig:couette-hex-unstruct}Cutaway of the
    unstructured hexahedral mesh with 1004 elements.}
\end{figure}

\paragraph{Three dimensional unstructured hexahedral mesh.}
As a further test a domain of dimension $[0,1]^3$ was considered.  This
domain was meshed using completely unstructured hexahedra.  Three levels
of refinement were used resulting in meshes with 96, 536 and 1004
elements.  A cutaway of the most refined mesh can be seen in
\autoref{fig:couette-hex-unstruct}.  Experimental $L^2$ errors and the
resulting orders of accuracy are presented in
\autoref{tab:couette-3d-us}.  Despite the fully unstructured nature of
the mesh the expected order of accuracy was again obtained in all cases.
We do, however, note the higher standard errors associated with these
results.

\begin{table}
  \centering
  \caption{\label{tab:couette-3d-us}$L^2$ energy errors and orders of
    accuracy for the Couette flow problem on three unstructured
    hexahedral meshes.  Mesh spacing was taken as $h \sim N^{-1/3}_E$
    where $N_E$ is the total number of elements in the mesh.}
  \begin{tabular}{rlll} \toprule
    & \multicolumn{3}{c}{$\sigma(t_\infty)\,/\,\si{\joule\per\metre\cubed}$} \\
    \cmidrule{2-4}
    \ccol{Hexes} & \ccol{$\wp=1$} & \ccol{$\wp = 2$} &
    \ccol{$\wp = 3$} \\
    \midrule
       96 & $1.91 \times 10^{1}$ & $4.32 \times 10^{-2}$ & $5.83 \times 10^{-4}$ \\
      536 & $8.20 \times 10^{0}$ & $9.11 \times 10^{-3}$ & $6.89 \times 10^{-5}$ \\
     1004 & $3.82 \times 10^{0}$ & $3.22 \times 10^{-3}$ & $2.04 \times 10^{-5}$ \\ \midrule
    \multicolumn{1}{r}{Order}
         & {$1.93 \pm 0.46$} & {$3.19 \pm 0.48$}
         & {$4.16 \pm 0.44$} \\ \bottomrule
  \end{tabular}
\end{table}

\subsection{Compressible Navier-Stokes Equations: Flow Over a Cylinder}

\begin{figure}
  \centering
  \includegraphics{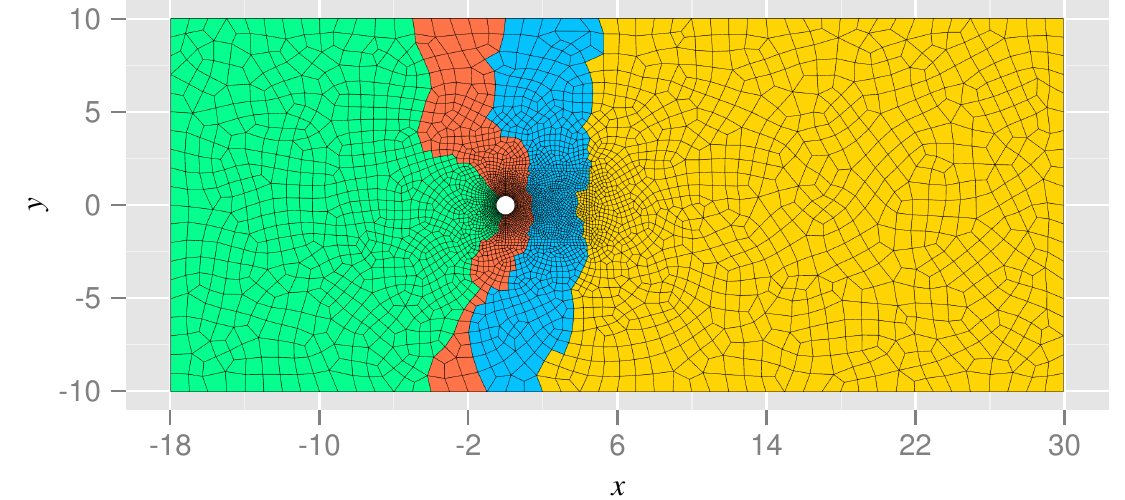}%
  \caption{\label{fig:cyl-partitions}Cross section in the $x$-$y$ plane
    of the cylinder mesh.  Colours indicate the partition to which the
    elements belong.}
\end{figure}

In order to demonstrate the ability of PyFR to solve the unsteady
Navier-Stokes equations flow over a cylinder at Reynolds number 3900 and
Mach number $M = 0.2$ was simulated.  A cylinder of radius $1/2$ was
placed at $(0,0)$ inside of a domain of dimension $[-18,30] \times
[-10,10] \times [0,3.2]$.  This domain was then meshed in the $x$-$y$
plane with 4661 quadratically curved quadrilateral elements.  Next, this
grid was extruded along the $z$-axis to yield a total of 46610
hexahedra.  The grid, which can be seen in \autoref{fig:cyl-partitions},
was partitioned into four pieces.  Along surfaces of $y = \pm 10$ and $x
= -18$ the inflow boundary condition of \autoref{sec:bc-sup-inflow} in
\autoref{sec:boundary-conditions} was imposed.  Along the surface of $x
= 30$ the outflow condition of \autoref{sec:bc-sub-outflow} in
\autoref{sec:boundary-conditions} was used.  Periodic conditions were
imposed in the $z$ direction.  On the surface of the cylinder the
no-slip isothermal wall condition of \autoref{sec:bc-no-slip-isot-wall}
in \autoref{sec:boundary-conditions} was imposed.  The free-stream
conditions were taken to be $\rho = 1$, $\vec{v} = \hat{\vec{x}}$, and
$p = 1/\gamma M^2$.  These were also used as the initial conditions for
the simulation.  DG correction functions were used with the LDG
parameters being $\beta = 1/2$ and $\tau = 0.1$.  The ratio of specific
heats was taken as $\gamma = 1.4$ and the Prandtl number as $P_r =
0.72$.

\begin{figure}
  \centering
  \includegraphics[width=\textwidth]{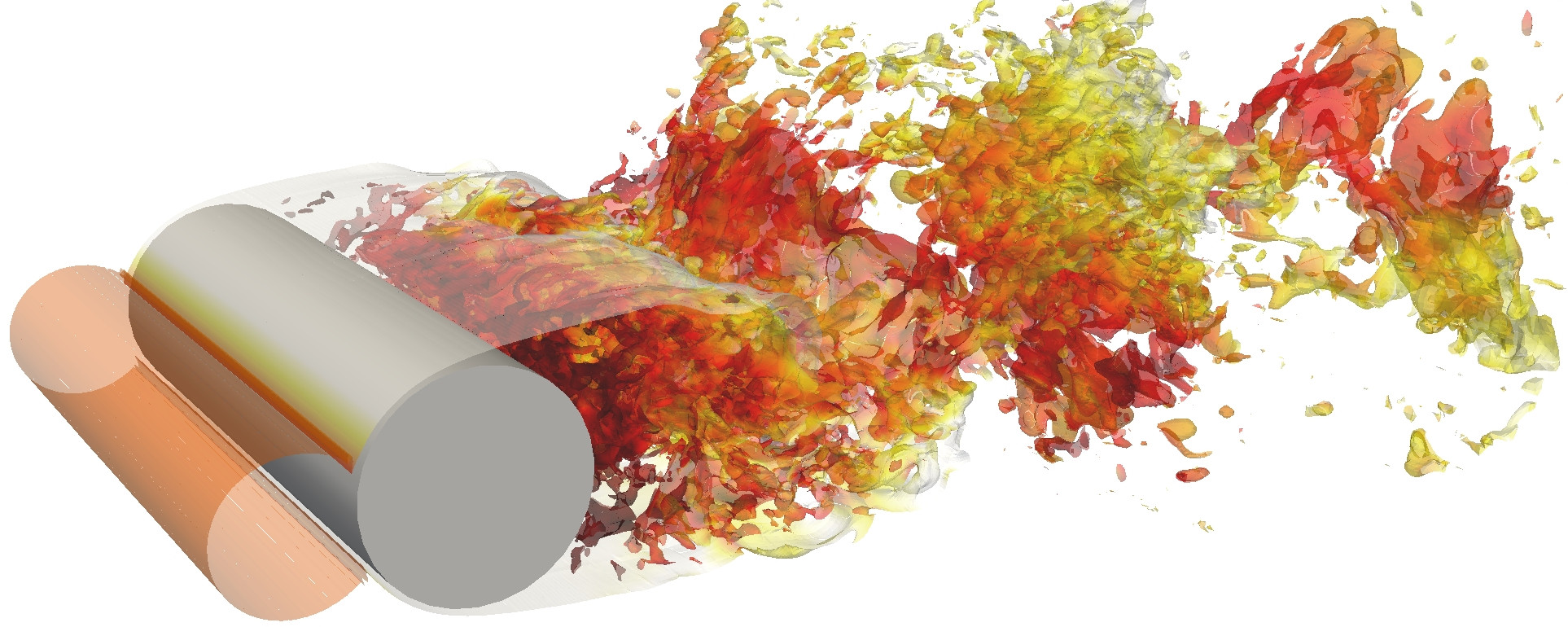}
  \caption{\label{fig:cyl-density}Isosurfaces of density around the
    cylinder.}
\end{figure}

The simulation was run with $\wp = 4$ with four NVIDIA K20c GPUs.  It
contained some $29 \times 10^6$ degrees of freedom.  Isosurfaces of
density captured after the turbulent wake had fully developed can be
seen in \autoref{fig:cyl-density}.

\section{Single Node Performance}
\label{sec:performance}

The single node performance of PyFR has been evaluated on an NVIDIA
M2090 GPU.  This accelerator has a theoretical peak double precision
floating point performance of $665\,\text{GFLOP/s}$, and when ECC is
disabled the theoretical peak memory bandwidth is $177\,\text{GB/s}$.
As points of reference we observe that cuBLAS (CUDA 5.5) is able to
obtain $407\,\text{GFLOP/s}$ when multiplying a pair of $4096\times4096$
matrices on this hardware, and the maximum device bandwidth obtainable
by the bandwidth test application included with the CUDA SDK is
$138.9\,\text{GiB/s}$ when ECC is disabled.  We shall refer to these
values as \emph{realisable peaks}.

To conduct the evaluation a fully periodic cuboidal domain was meshed
with $50\,176$ hexahedral elements.  The double precision Navier-Stokes
solver of PyFR was then run on this mesh at orders $\wp = 2,3,4$ with
$\beta = 1/2$.  In conducting the analysis kernels were grouped into one
of three categories: matrix multiplications (DGEMM), point-wise kernels
with direct memory access patterns (PD) and point-wise kernels with some
level of indirect memory access (PI). Indirection arises in the
computation of $\mathfrak{C}_{\alpha}$ in \autoref{eq:commsoln} and
$\mathfrak{F}_{\alpha}$ in \autoref{eq:commpflux} and occurs as a
consequence of the unstructured nature of PyFR.  The resulting
breakdowns of wall-clock time, memory bandwidth and floating point
operations can be seen in \autoref{tab:single-node}.  It is clear that he
majority of floating point operations are concentrated inside the calls
to DGEMM with the point-wise operations are heavily memory bandwidth
bound.  Of this bandwidth some ${\sim}15\%$ was ascribed to register
spillage above and beyond that which can be absorbed by the L1 cache.

\begin{table}
  \centering
  \caption{\label{tab:single-node}Single GPU performance of PyFR for the
    Navier-Stokes equations when run on an NVIDIA M2090 with ECC
    disabled.  As the memory bandwidth requirements of DGEMM are
    dependent on the accumulation strategy adopted by the implementation
    these values have been omitted.}
  \begin{tabular}{rrrrr} \toprule
    & & \multicolumn{3}{c}{Order} \\
    \cmidrule{3-5}
    & & \ccol{$\wp = 2$} & \ccol{$\wp = 3$} & \ccol{$\wp = 4$} \\ \midrule
    Wall time / \% \\
    & DGEMM & $55.7$ & $66.2$ & $81.4$ \\
    & PD    & $24.9$ & $21.5$ & $12.8$ \\
    & PI    & $19.4$ & $12.3$ & $5.8$ \\ \midrule
    Bandwidth / $\text{GiB/s}$ \\
    & PD   & $125.5$ & $125.0$ & $124.8$ \\
    & PI    & $124.8$ & $124.3$ & $124.2$ \\ \midrule
    Arithmetic / $\text{GFLOP/s}$ \\
    & DGEMM & $205.3$ & $368.1$ & $305.4$ \\
    & PD    & $0.7$   & $0.7$   & $0.7$ \\
    & PI    & $0.9$   & $0.8$   & $0.9$ \\ \bottomrule
  \end{tabular}
\end{table}

The high fraction of peak bandwidth obtained by the indirect
kernels can be attributed to three factors.  Firstly, the constant data required for calculations at ????, such as
$\hat{\vec{n}}^{(f)}_{e\sigma n}$ and $J^{(f)}_{e\sigma
  n}n^{(f)}_{e\sigma n}$, is ordered to ensure direct (coalesced)
access.  Secondly, at start-up PyFR attempts to determine an iteration
ordering over the various flux-point pairs that will minimise the number
of cache misses.

Many of the memory accesses are therefore are
near-coalesced.  Thirdly and finally we highlight the impressive
latency-hiding capabilities of the CUDA programming model.

In line with expectations the proportion of time spent performing
matrix-matrix multiplications increases as a function of order.  When
going from $\wp = 2$ to $\wp = 3$ a significant portion of the
additional compute is offset by the improved performance of cuBLAS.
However, when $\wp = 4$ the performance of these kernels in absolute
terms can be seen to regress slightly.  This contributes to the greatly
increased fraction of wall-clock time spent inside of these kernels.
Nevertheless, the achieved rate of $305.4 \text{GFLOP/s}$ is still over
$75\%$ of the realisable peak.  Also in line with expectations is
the invariance of the arithmetic performance of the point-wise kernels
with respect to order.  As the order is varied all that changes is the
number of points to be processed with the operation itself remaining
identical.

\section{Scalability}
\label{sec:scalability}

The scalability of PyFR has been evaluated on the Emerald GPU cluster.
It is housed at the STFC Rutherford Appleton Laboratory and based around
60 HP SL390 nodes with three NVIDIA M2090 GPUs and 24 HP SL390 nodes
with eight NVIDIA M2090 GPUs.  Nodes are connected by QDR InfiniBand.

For simplicity all runs herein were performed on the eight GPU nodes.
As a starting point a domain of dimension $[-16,16] \times [-16,16]
\times [0,1.75]$ was meshed isotropically with $N_E = 114\,688$ structured
hexahedral elements.  The mesh was configured with completely periodic
boundary conditions.  When run with the Navier-Stokes solver in PyFR
with $\wp = 3$ the mesh gives a working set of
${\sim}4720\,\text{MiB}$.  This is sufficient to 90\% load an M2090
which when ECC is enabled has ${\sim}5250\,\text{MiB}$ memory
available to the user.  When examining the scalability of a code there
are two commonly used metrics.  The first of these is weak scalability
in which the size of the target problem is increased in proportion to
the number of ranks $N$ with $N_E \propto N$.  For a code with perfect
weak scalability the runtime should remain unchanged as more ranks are
added.  The second metric is strong scalability wherein the problem size
is fixed and the speedup compared to a single rank is assessed.  Perfect
strong scalability implies that the runtime scales as $1/N$.

For the domain outlined above weak scalability was evaluated by
increasing the dimensions of the domain according to $[-16,16] \times
N[-16,16] \times [0,1.75]$.  This extension permitted the domain to be
trivially decomposed along the $y$-axis.  The resulting runtimes for $1
\le N \le 104$ can be seen in \autoref{tab:ns-weak-scale}.  We note that
in the $N = 104$ case that the simulation consisted of some $3.8 \times
10^9$ degrees of freedom with a working set of
${\sim}485\,\text{GiB}$.

\begin{table}
  \centering
  \caption{\label{tab:ns-weak-scale}Weak scalability of PyFR for the
    Navier-Stokes equations with $\wp = 3$.  Runtime is normalised with
    respect to a single NVIDIA M2090 GPU.}
  \begin{tabular}{r|llllllll} \toprule
    \# M2090s & 1    & 2    & 4    & 8    & 16   & 32   & 64   & 104 \\
    Runtime   & 1.00 & 1.00 & 1.01 & 1.01 & 1.01 & 1.01 & 1.01 & 1.01 \\
    \bottomrule
  \end{tabular}
\end{table}

To study the strong scalability the initial domain was partitioned along
the $x$- and $y$-axes.  Each partition contained exactly $N_E/N$s.  The
resulting speedups for $1 \le N \le 32$ can be seen in
\autoref{tab:ns-strong-scale}.  Up to eight GPUs scalability can be seen
to be near perfect.  Beyond this the relationship begins to break down.
When $N = 32$ an improvement of 26 can be observed.  However, in this
case each GPU is loaded to less than 3\% and so the result is to be
expected.

\begin{table}
  \centering
  \caption{\label{tab:ns-strong-scale}Strong scalability of PyFR for the
    Navier-Stokes equations with $\wp = 3$.  The speedup is relative to
    a single NVIDIA M2090 GPU.}
  \begin{tabular}{r|llllll} \toprule
    \# M2090s & 1    & 2    & 4    & 8    & 16   & 32 \\
    Speedup   & 1.00 & 2.03 & 3.96 & 7.48 & 14.07 & 26.18 \\
    \bottomrule
  \end{tabular}
\end{table}

\section{Conclusions}
\label{sec:conclusion}

In this paper we have described PyFR, an open source Python based
framework for solving advection-diffusion type problems on streaming
architectures.  The structure and ethos of PyFR has been explained
including our methodology for targeting multiple hardware platforms.  We
have shown that PyFR exhibits spatial super accuracy when solving the 2D
Euler equations and the expected order of accuracy when solving Couette
flow problem on a range of grids in 2D and 3D.  Qualitative results for
unsteady 3D viscous flow problems on curved grids have also been
presented.  Performance of PyFR has been validated on an NVIDIA M2090
GPU in three dimensions.  It has been shown that the compute bound
kernels are able to obtain between $50\%$ and $90\%$ of realisable peak
FLOP/s whereas the bandwidth bound point-wise kernels are, across the
board, able to obtain in excess of $89\%$ realisable peak bandwidth.
The scalability of PyFR has been demonstrated in the strong sense up to
32 NVIDIA M2090s and in the weak sense up to 104 NVIDIA M2090s when
solving the 3D Navier-Stokes equations.

\section*{Acknowledgements}
The authors would like to thank the Engineering and Physical Sciences
Research Council for their support via two Doctoral Training Grants and
an Early Career Fellowship (EP/K027379/1). The authors would also like
to thank the e-Infrastructure South Centre for Innovation for granting
access to the Emerald supercomputer, and NVIDIA for donation of three
K20c GPUs.

\appendix

\section{Matrix Representation}
\label{sec:mat-rep}

It is possible to cast the majority of operations in an FR step as
matrix-matrix multiplications of the form
\begin{equation}
  \mat{C} \leftarrow c_1\mat{A}\mat{B} + c_2\mat{C},
\end{equation}
where $c_{1,2} \in \mathbb{R}$ are constants, $\mat{A}$ is a constant
operator matrix, and $\mat{B}$ and $\mat{C}$ are state matrices.  To accomplish this we
start by introducing the following constant operator matrix
\begin{align*}
 \big(\mat{M}^{0}_{e}\big)_{\sigma\rho} &= \ell^{(u)}_{e\rho}(\tilde{\vec{x}}^{(f)}_{e\sigma}),   & \dim \mat{M}^{0}_e &= N_e^{(f)} \times N_e^{(u)},
\end{align*}
and the following state matrices
\begin{align*}
 \big(\mat{U}^{(u)}_{e}\big)_{\rho(n\alpha)} &= u^{(u)}_{e\rho n\alpha}, & \dim \mat{U}^{(u)}_e &= N_e^{(u)} \times N_V|\vec{\Omega}_e|,\\
 \big(\mat{U}^{(f)}_{e}\big)_{\sigma(n\alpha)} &= u^{(f)}_{e\sigma n\alpha},     & \dim \mat{U}^{(f)}_e &= N_e^{(f)} \times N_V|\vec{\Omega}_e|.
\end{align*}
In specifying the state matrices there is a degree of freedom associated
with how the $N_V$ field variables for each element are packed along a row of the matrix,
with the possible packing choices being discussed in \autoref{sec:layout}.
Using these matrices we are able to reformulate \autoref{eq:ufpts} as
\begin{equation} \label{eq:ufptsm}
  \mat{U}^{(f)}_{e} = \mat{M}^{0}_{e}\mat{U}^{(u)}_{e}.
\end{equation}
In order to apply a similar procedure to \autoref{eq:transqu} we let
\begin{align*}
  \big(\mat{M}^{4}_{e}\big)_{\rho\sigma} &=
  \big[\tilde{\grad}\ell^{(u)}_{e\rho}(\tilde{\vec{x}})\big]_{\tilde{\vec{x}}
    = \tilde{\vec{x}}^{(u)}_{e\sigma}}, & \dim \mat{M}^{4}_{e},
  &= N_{D}N_{e}^{(u)} \times N_{e}^{(u)},\\
  \big(\mat{M}^{6}_{e}\big)_{\rho\sigma} &=
  \big[\hat{\tilde{\vec{n}}}^{(f)}_{e\rho}\cdot \tilde{\grad} \cdot
  \vec{g}^{(f)}_{e\rho}(\tilde{\vec{x}})\big]_{\tilde{\vec{x}} =
    \tilde{\vec{x}}^{(f)}_{e\sigma}}, & \dim \mat{M}^{6}_{e},
  &= N_{D}N_{e}^{(u)} \times N_{e}^{f},\\
 \big(\mat{C}^{(f)}_{e}\big)_{\rho(n\alpha)} &=
  \mathfrak{C}_{\alpha}u^{(f)}_{e\rho n\alpha},
  & \dim \mat{C}^{(f)}_{e} &= N^{(f)}_{e} \times
  N_V\abs{\vec{\Omega}_e},\\
  \big(\tilde{\mat{Q}}^{(u)}_{e}\big)_{\sigma(n\alpha)} &=
  \tilde{\vec{q}}^{(u)}_{e\sigma n\alpha}, &
  \dim \tilde{\mat{Q}}^{(u)}_{e}  &= N_DN^{(u)}_{e} \times
  N_V\abs{\vec{\Omega}_e},
\end{align*}
Here it is important to qualify assignments of the form $\mat{A}_{ij} =
\vec{x}$ where $\vec{x}$ is a $N_D$ component vector.  As above there is
a degree of freedom associated with the packing.  With the benefit of
foresight we take the stride between subsequent elements of $\vec{x}$ in
a matrix column to be either $\Delta i = N^{(u)}_{e}$ or $\Delta i =
N^{(f)}_{e}$ depending on the context.  With these matrices
\autoref{eq:transqu} reduces to
\begin{equation} \label{eq:tquptsm}
\begin{aligned}
  \tilde{\mat{Q}}^{(u)}_{e} &= \mat{M}^{6}_{e}\big\{\mat{C}^{(f)}_{e} -
  \mat{U}^{(f)}_{e}\big\} + \mat{M}^{4}_{e}\mat{U}^{(u)}_{e}\\
  &= \mat{M}^{6}_{e}\big\{\mat{C}^{(f)}_{e} -
  \mat{M}^{0}_{e}\mat{U}^{(u)}_{e}\big\} +
  \mat{M}^{4}_{e}\mat{U}^{(u)}_{e}\\
  &= \mat{M}^{6}_{e}\mat{C}^{(f)}_{e} +
  \big\{\mat{M}^{4}_{e} - \mat{M}^{6}_{e}\mat{M}^{0}_{e}\big\}\mat{U}^{(u)}_{e}.
\end{aligned}
\end{equation}
Applying the procedure to \autoref{eq:gradufpts} we take
\begin{align*}
  \mat{M}^{5}_{e} &= \diag(\mat{M}^{0}_{e}, \ldots, \mat{M}^{0}_{e}) &
  \dim{\mat{M}^{5}_{e}} &= N_DN^{(f)}_{e} \times N_DN^{(u)}_{e},\\
  \big(\mat{Q}^{(u)}_{e}\big)_{\sigma(n\alpha)} &=
  \vec{q}^{(u)}_{e\sigma n\alpha}, &
  \dim{\mat{Q}^{(u)}_{e}} &= N_DN^{(u)}_{e} \times
  N_V\abs{\vec{\Omega}_e},\\
  \big(\mat{Q}^{(f)}_{e}\big)_{\sigma(n\alpha)} &=
  \vec{q}^{(f)}_{e\sigma n\alpha}, &
  \dim{\mat{Q}^{(f)}_{e}} &= N_DN^{(f)}_{e} \times
  N_V\abs{\vec{\Omega}_e},
\end{align*}
hence
\begin{equation}
  \mat{Q}^{(f)}_{e} = \mat{M}^{5}_{e}\mat{Q}^{(u)}_{e},
\end{equation}
where we note the block diagonal structure of $\mat{M}^{5}_{e}$.  This
is a direct consequence of the above choices for $\Delta i$.  Finally, to
rewrite \autoref{eq:tdivflux} we write
\begin{align*}
  \big(\mat{M}^{1}_{e}\big)_{\rho\sigma} &=
  \big[\tilde{\grad}\ell^{(u)}_{e\rho}(\tilde{\vec{x}})\big]^T_{\tilde{\vec{x}}
    = \tilde{\vec{x}}^{(u)}_{e\sigma}}, & \dim \mat{M}^{1}_{e} &=
  N^{(u)}_{e} \times N_DN^{(u)}_{e},\\
  \big(\mat{M}^{2}_{e}\big)_{\rho\sigma} &=
  \big[\ell^{(u)}_{e\rho}(\tilde{\vec{x}}^{(f)}_{e\sigma})
  \hat{\tilde{\vec{n}}}^{(f)}_{e\sigma}\big]^T, & \dim \mat{M}^{2}_{e} &=
  N^{(f)}_{e} \times N_DN^{(u)}_{e},\\
  \big(\mat{M}^{3}_{e}\big)_{\rho\sigma} &= \big[\tilde{\grad} \cdot
  \vec{g}^{(f)}_{e\sigma}(\tilde{\vec{x}})\big]_{\tilde{\vec{x}} =
    \tilde{\vec{x}}^{(u)}_{e\rho}}, & \dim \mat{M}^{3}_{e} &=
  N^{(u)}_{e}
  \times N^{(f)}_{e},\\
  \big(\tilde{\mat{D}}^{(f)}_{e}\big)_{\sigma(n\alpha)} &=
  \mathfrak{F}^{\vphantom{(f_\perp)}}_{\alpha}\tilde{f}^{(f_\perp)}_{e\sigma
    n\alpha}, & \dim \tilde{\mat{D}}^{(f)}_{e} &= N^{(f)}_{e} \times N_V\abs{\vec{\Omega}_e},\\
  \big(\tilde{\mat{F}}^{(u)}_{e}\big)_{\rho(n\alpha)} &=
  \tilde{\vec{f}}^{(u)}_{e\rho n\alpha}, & \dim
  \tilde{\mat{F}}^{(u)}_{e} &= N_DN^{(u)}_{e} \times
  N_V\abs{\vec{\Omega}_e},\\
  \big(\tilde{\mat{R}}^{(u)}_{e}\big)_{\rho(n\alpha)} &= (\tilde{\grad}
  \cdot \tilde{\vec{f}})^{(u)}_{e\rho n\alpha}, & \dim
  \tilde{\mat{R}}^{(u)}_{e} &= N_e^{(u)} \times N_V\abs{\vec{\Omega}_e},
\end{align*}
and after substitution of \autoref{eq:tnormfluxf} for
$\tilde{f}^{(f_\perp)}_{e\sigma n\alpha}$ obtain
\begin{equation}
\begin{aligned}
  \tilde{\mat{R}}^{(u)}_{e} &=
  \mat{M}^{3}_{e}\big\{\tilde{\mat{D}}^{(f)}_{e} -
  \mat{M}^{2}_{e}\tilde{\mat{F}}^{(u)}_{e}\big\} +
  \mat{M}^{1}_{e}\tilde{\mat{F}}^{(u)}_{e}\\
  &= \mat{M}^{3}_{e}\tilde{\mat{D}}^{(f)}_{e} +
  \big\{\mat{M}^{1}_{e} - \mat{M}^{3}_{e}\mat{M}^{2}_{e}\big\}
  \tilde{\mat{F}}^{(u)}_{e}.
\end{aligned}
\end{equation}

\section{Approximate Riemann Solvers}
\label{sec:riemann-solvers}

\subsection{Overview}

In the following section we take $u_L$ and $u_R$ to be the two
discontinuous solution states at an interface and $\hat{\vec{n}}_L$ to
be the normal vector associated with the first state.  For convenience
we take $\vec{f}^{(\text{inv})}_L =
\vec{f}^{(\text{inv})}_{\vphantom{L}}(u^{\vphantom{(\text{inv})}}_L)$,
and $\vec{f}^{(\text{inv})}_R =
\vec{f}^{(\text{inv})}_{\vphantom{R}}(u^{\vphantom{(\text{inv})}}_R)$
with inviscid fluxes being prescribed by \autoref{eq:eulerflux}.

\subsection{Rusanov}

Also known as the local Lax-Friedrichs method a Rusanov type Riemann
solver imposes inviscid numerical interface fluxes according to
\begin{equation}
  \mathfrak{F}^{(\text{inv})} = \frac{\hat{\vec{n}}_L}{2} \cdot \left\{\vec{f}^{(\text{inv})}_L +
    \vec{f}^{(\text{inv})}_R\right\} +
  \frac{s}{2}(u_L - u_R),
\end{equation}
where $s$ is an estimate of the maximum wave speed
\begin{equation}
  s = \sqrt{\frac{\gamma(p_L + p_R)}{\rho_L + \rho_R}} +
  \frac{1}{2}\big|\hat{\vec{n}}_L \cdot (\vec{v}_L + \vec{v}_R)\big|.
\end{equation}

\section{Boundary Conditions}
\label{sec:boundary-conditions}

\subsection{Overview}

To incorporate boundary conditions into the FR approach we introduce a
set of boundary interface types $b \in \mathcal{B}$.  At a boundary
interface there is only a single flux point: that which belongs to the
element whose edge/face is on the boundary.  Associated with each
boundary type are a pair of functions $\mathfrak{C}^{(b)}_{\alpha}(u_L)$
and $\mathfrak{F}^{(b)}_{\alpha}(u_L,\vec{q}_L,\hat{\vec{n}}_L)$ where
$u_L$, $\vec{q}_L$, and $\hat{\vec{n}}_L$ are the solution, solution
gradient and unit normals at the relevant flux point.  These functions
prescribe the common solutions and normal fluxes, respectively.

Instead of directly imposing solutions and normal fluxes it is
oftentimes more convenient for a boundary to instead provide ghost
states.  In its simplest formulation $\mathfrak{C}^{(b)}_{\alpha} =
\mathfrak{C}_{\alpha}(u_L,\mathfrak{B}^{(b)} u_L)$ and
$\mathfrak{F}^{(b)}_{\alpha} =
\mathfrak{F}_{\alpha}(u_L,\mathfrak{B}^{(b)} u_L,\vec{q}_L,
\mathfrak{B}^{(b)} \vec{q}_L, \hat{\vec{n}}_L)$ where
$\mathfrak{B}^{(b)} u_L$ is the ghost solution state and
$\mathfrak{B}^{(b)} \vec{q}_L$ is the ghost solution gradient.  It is
straightforward to extend this prescription to allow for the
provisioning of different ghost solution states for
$\mathfrak{C}_{\alpha}$ and $\mathfrak{F}_{\alpha}$ and to permit
$\mathfrak{B}^{(b)} \vec{q}_L$ to be a function of $u_L$ in addition to
$\vec{q}_L$.

\subsection{Supersonic Inflow}
\label{sec:bc-sup-inflow}

The supersonic inflow condition is parameterised by a free-stream
density $\rho_f$, velocity $\vec{v}_f$, and pressure $p_f$.

\begin{align}
  \mathcal{B}^{(\text{inv})} u_L = \mathcal{B}^{(\text{ldg})} u_L &=
  \begin{Bmatrix}
    \rho_f\\
    \rho_f\vec{v}_f\\
    p_f/(\gamma - 1) + \frac{\rho_f}{2}\|\vec{v}_f\|^2
  \end{Bmatrix},\\
  \mathcal{B}^{(\text{ldg})} \vec{q}_L &= 0,
\end{align}

\subsection{Subsonic Outflow}
\label{sec:bc-sub-outflow}

Subsonic outflow boundaries are parameterised by a free-stream pressure
$p_f$.

\begin{align}
  \mathcal{B}^{(\text{inv})} u_L = \mathcal{B}^{(\text{ldg})} u_L &=
  \begin{Bmatrix}
    \rho_L\\
    \rho_L\vec{v}_L\\
    p_f/(\gamma - 1) + \frac{\rho_L}{2}\|\vec{v}_L\|^2
  \end{Bmatrix},\\
  \mathcal{B}^{(\text{ldg})} \vec{q}_L &= 0,
\end{align}

\subsection{No-slip Isothermal Wall}
\label{sec:bc-no-slip-isot-wall}

The no-slip isothermal wall condition depends on the wall temperature
$C_pT_w$ and the wall velocity $\vec{v}_w$.  Usually $\vec{v}_w = 0$.

\begin{align}
  \mathcal{B}^{(\text{inv})} u_L &= \rho_L\begin{Bmatrix}
    1\\
    2\vec{v}_w -\vec{v}_L \\
    C_pT_w/\gamma + \frac{1}{2}\norm{2\vec{v}_w - \vec{v}_L}^2
  \end{Bmatrix},\\
  \mathcal{B}^{(\text{ldg})} u_L &= \rho_L\begin{Bmatrix}
    1\\
    \vec{v}_w\\
    C_pT_w/\gamma + \frac{1}{2}\norm{\vec{v}_w}^2
  \end{Bmatrix},\\
  \mathcal{B}^{(\text{ldg})} \vec{q}_L &= \vec{q}_L,
\end{align}

\end{document}